%% file: manuscript.tex
\pdfoutput=1

\documentclass[aps,pra,reprint,superscriptaddress,nofootinbib,floatfix,
  showkeys]{revtex4-2}

\usepackage{amsmath,amssymb,bm}
\usepackage{graphicx}
\usepackage{microtype}
\usepackage{xcolor}
\usepackage[colorlinks=true,linkcolor=blue,citecolor=blue,urlcolor=blue]{hyperref}
\hypersetup{
  pdftitle={clifford\_qc: A Python Toolkit for Quantum Simulation},
  pdfauthor={Ginanjar Utama and Hermawan Kresno Dipojono},
  pdfsubject={Software architecture, evidence labelling, admissibility and
    resource lifetimes for operator-centric quantum simulation}
}

\newcommand{\ket}[1]{\lvert #1\rangle}
\newcommand{\avg}[1]{\langle #1\rangle}

\newcommand{\pkg}{\texttt{clifford\_qc}}
\newcommand{\Ctime}{C_{\mathrm{time}}(\epsilon)}

\newcommand{\tightcolumns}{\setlength{\tabcolsep}{4pt}}
\newcommand{\compactcolumns}{\setlength{\tabcolsep}{2.5pt}}

\input{tables/numbers.tex}

\begin{document}

\title{\pkg{}: A Python Toolkit for Quantum Simulation}

\author{Ginanjar Utama}
\email{ginanjar.utama@gmail.com}
\affiliation{Department of Engineering Physics, Institut Teknologi Bandung,
Bandung, Indonesia}

\author{Hermawan Kresno Dipojono}
\email{dipojono@itb.ac.id}
\affiliation{Department of Engineering Physics, Institut Teknologi Bandung,
Bandung, Indonesia}

\date{September 19, 2026; revised November 1, 2026}

\begin{abstract}
\pkg{} is a Python research toolkit for constructing quantum models, comparing
variational and subspace eigensolvers, and estimating measurement resources.
A common sparse Pauli algebra connects Hamiltonians, density operators, gates,
and candidate-selection observables; dedicated interfaces handle circuits,
execution, and statistical estimates. A worked gradient example illustrates
shared measurements, and a molecular workflow separates reusable preparation,
solving, and reference validation. Matrix-element banks support virtual
subspaces, exact Ritz residuals, and folded-spectrum roots. Process-isolated
benchmarks using the legacy density reference show that packed coefficients and
streaming reduce observed peak memory on the larger tested inputs at increased
runtime. Measurement studies
separate setting counts, shot requirements, and device cost: fully commuting
readout certifies a fixed first-order Ritz functional with fewer shots, while
anticommuting cliques help some fixed reference energies and not others.
Entangling-gate time and connectivity can offset shot savings or make a
protocol inadmissible under illustrative device models. Further studies expose
subspace bias, indeterminate encoding comparisons, failed approximate-restriction
screens, and high residual-measurement costs under the tested uniform
allocations. A provider workflow, scoped cost ledger, portable evidence export, and
independent checks connect calculations to their recorded assumptions. Generated
tables, provenance records, and manuscript checks connect claims to inputs. These
small-system results support reproducible method development and establish
neither hardware performance nor a general simulation speedup.
\end{abstract}


\maketitle

\section{Introduction}
\label{sec:introduction}

Developing a quantum simulation method requires more than an energy estimate.
Researchers need to construct a Hamiltonian, choose a reference state, compare
algorithms, account for measurements, and check results against a classical
calculation. These steps often use different representations and independently
written bookkeeping. A shared implementation helps make comparisons consistent
and lets a useful component be reused outside the experiment that introduced it.

\pkg{} provides that implementation around sparse Pauli-word operators. Its
main use is small-system research on quantum algorithms for spin, lattice, and
molecular models. Users can work directly with the operator algebra, execute a
circuit program, or build a projected eigensolver on a common model. The same
measurement data can support several observables and projected matrix elements.
Explicit storage policies and reusable model preparation address the classical
cost of repeated experiments.

Subspace methods are a particular focus. Quantum subspace expansion,
nonorthogonal variational methods, and Krylov methods reconstruct a small
eigenproblem from expectation values
\cite{mcclean2017qse,huggins2020nonorthogonal,stair2020krylov}.
Their accuracy depends on both the chosen subspace and the stability of its
overlap matrix. Finite sampling can amplify errors in a poorly conditioned
basis \cite{epperly2022theory,lee2024sampling,zhang2024measurement}.
Consequently, the package records basis size, conditioning, measurement support,
and uncertainty alongside energy, and on request the exact residual that tells
how far a Ritz state is from an eigenstate. The underlying operator formulation
is developed in a companion manuscript \cite{utama2026operatorcentric};
Sec.~\ref{sec:operator} introduces the part of it a user of the software meets,
and the rest of this paper explains the software that supports it and the other
implemented methods.

The contribution is the integration of reusable components: one operator
algebra, explicit execution interfaces, shared measurement data, exact residual
diagnostics, configurable coefficient storage, and benchmark records with stated
assumptions.

\pkg{} is not a replacement for the established quantum software stack; it is
built to sit beside it. General-purpose frameworks such as Qiskit
\cite{javadiabhari2024qiskit} and PennyLane \cite{bergholm2018pennylane} cover
circuit construction, differentiation, and hardware execution; OpenFermion
\cite{mcclean2020openfermion} and Tequila \cite{kottmann2021tequila} cover
electronic-structure workflows and algorithm prototyping; pytket
\cite{sivarajah2020tket} compiles and routes, PyZX \cite{kissinger2020pyzx}
rewrites circuit diagrams, and Stim \cite{gidney2021stim} specializes in
stabilizer simulation. \pkg{} occupies a narrower position: one sparse Pauli
algebra connecting the operators these tasks use; measurement data
reused across observables and projected matrix elements; and result records
that carry the evidence category and the device assumptions behind each number.
Optional bridges keep the other tools reachable from the same model. We make no
advantage claim over those packages and provide no comparative benchmark
against them.

PauliEngine \cite{muller2026pauliengine} is a particularly relevant
operator-level comparator. Its C++ backend with Python bindings uses bitwise
Pauli arithmetic and supports both numerical and SymEngine symbolic coefficients,
including symbolic differentiation and substitution. Its authors benchmark
Hamiltonian multiplication against OpenFermion and PauliArray, and dynamical
Lie-algebra closure and structure-constant construction against PennyLane.
Their reported timings favor PauliEngine in the larger tested multiplication
regimes, while PauliArray is faster on some smaller inputs; the Lie-algebra
results concern the specific tested families and hardware. Those results do not
benchmark \pkg{} and cannot establish a speed ratio between the packages.
The PauliEngine source is at \url{https://github.com/tequilahub/pauliengine}.

Both tools manipulate sparse Pauli operators, but their principal boundaries
differ. PauliEngine emphasizes optimized operator arithmetic and symbolic
coefficients; \pkg{} uses numerical complex coefficients in its Pauli kernel
and connects them to reference providers, adaptive projected eigensolvers,
measurement analysis, and evidence workflows. The comparison is one of
capability and experimental scope, not overall performance. Using PauliEngine
as a backend would require an explicit adapter and checks for qubit ordering,
phase conventions, coefficient and tolerance semantics, and conversion cost;
no such integration or matched benchmark is claimed here. A future comparison
should use identical operators and outputs, check products and commutators
independently, and report conversion, peak memory, and end-to-end workflow cost
as well as kernel time.

Section~\ref{sec:operator} introduces the operator-centric representation and
works a small example through it.
Section~\ref{sec:features} introduces the capabilities, a typical workflow with
its residual diagnostics, and a molecular run through the pipeline.
Section~\ref{sec:architecture} explains how that representation is implemented
and what the interfaces are.
Section~\ref{sec:scopes} describes uncertainty, device cost, and storage limits.
Sections~\ref{sec:gates}--\ref{sec:refused} connect these mechanisms to the
committed evidence. The final sections discuss limitations and reproducibility.

\section{The operator-centric representation}
\label{sec:operator}

The organizing principle is a common algebra for the operators a calculation
uses. Density operators, gates, observables, and candidate-selection
observables can share multiplication, adjoints, and expectation-value
evaluation. This continues the use of Clifford algebra for quantum
operations \cite{somaroo1998operations}; the companion manuscript develops
the package's conventions and algorithmic formulation
\cite{utama2026operatorcentric}. Here we explain the representation through
the interfaces and a small example.

\subsection{A common operator algebra}
\label{sec:operator-algebra}

An $n$-qubit operator is expanded in the Hermitian Pauli-word basis
$\mathcal{P}_n=\{I,X,Y,Z\}^{\otimes n}$,
\begin{equation}
 A=\sum_{W\in\mathcal{P}_n}a_W W.
 \label{eq:expansion}
\end{equation}
This gives a realization of the complex Clifford algebra
$\mathrm{Cl}(2n,\mathbb{C})\cong M(2^n,\mathbb{C})$
\cite{hrdina2022clifford,silva2025operator}. Jordan--Wigner maps fermionic
operators to Pauli strings in this algebra \cite{jordan1928}; parity and
Bravyi--Kitaev encodings are also available
\cite{bravyi2002fermionic,seeley2012bravyikitaev}. The isomorphism gives the
same operator space as matrix mechanics. It introduces no extra expressivity
and makes no general promise of sparse storage.

The kernel stores Eq.~\eqref{eq:expansion} in a multivector, \texttt{MV},
mapping packed word codes to complex coefficients. If
$\rho=\sum_W r_W W$, an expectation value is the trace pairing
\begin{equation}
 \mathrm{Tr}(A\rho)=2^n\sum_W a_W r_W.
 \label{eq:trace-pairing}
\end{equation}
Only shared support contributes. A gate acts by $U\rho U^\dagger$; a channel
acts by a sum of $K_a\rho K_a^\dagger$ over its Kraus operators. Each $K_a$
can be an \texttt{MV}, while the channel itself is a map. Density operators
and unitaries have different physical constraints despite sharing an
algebraic container.

The software retains separate types where they serve different tasks.
\texttt{PauliSum} is the public Hamiltonian and observable container;
\texttt{Program} holds circuit instructions; statevector and stabilizer
backends keep their own state representations. Samples, estimates, and
uncertainty reports are separate data objects, not multivectors. Shared
Pauli codes and explicit conversions connect these interfaces.
The practical benefit is reuse of operator products and measurement
functionals without introducing another operator basis at each layer.

\subsection{A worked example: gradients and shared measurements}
\label{sec:operator-example}

Two Pauli words multiply to one Pauli word times a phase. The packed kernel
computes the word code by exclusive-or and the phase by bit-mask parity
operations. A pair therefore either commutes or anticommutes, so a
commutator can be assembled directly from the anticommuting Hamiltonian
terms.

Consider the two-site transverse-field Ising Hamiltonian, with couplings
set to unity,
\begin{equation}
 H=Z_pZ_q+X_p+X_q.
 \label{eq:tfim}
\end{equation}
For a fixed reference $\rho$, append
$R_P(\theta)=\exp(-\mathrm{i}\theta P/2)$ and define
$E_P(\theta)=\mathrm{Tr}(H R_P(\theta)\rho R_P^\dagger(\theta))$.
Its initial slope is
\begin{equation}
 g_P=E_P'(0)=\mathrm{Tr}(G_P\rho),\qquad
 G_P=-\frac{\mathrm{i}}{2}[H,P].
 \label{eq:score}
\end{equation}
This is the normalization used by \texttt{CommutatorBank}. For three
candidate words, direct multiplication gives
\begin{align}
 G_{Y_p} &= Z_p-X_pZ_q, \nonumber\\
 G_{Y_pZ_q} &= Z_pZ_q-X_p-Y_pY_q, \label{eq:gradients}\\
 G_{Z_pY_q} &= Z_pZ_q-X_q-Y_pY_q. \nonumber
\end{align}
The expressions hold for any reference; the reference determines their
expectation values and hence the ranking by $|g_P|$.

This example separates new measurement work from reuse. The energy already
requires $Z_pZ_q$, $X_p$, and $X_q$. The candidate scores additionally require
$Z_p$, $X_pZ_q$, and $Y_pY_q$: response words arise from products of
Hamiltonian terms with candidate words, and need not occur in the
Hamiltonian itself. Once estimated, the means of $Z_pZ_q$ and $Y_pY_q$
contribute to both of the last two scores without separate samples for
each occurrence. The shared cache exposes this overlap; it does not make
the additional observables free.

Reuse is conditional on the state and measurement plan. For a fixed
reference and compatible readout, the same outcomes can feed several
linear functionals, with within-setting covariance retained by the grouped
estimator. An ADAPT circuit update changes the reference and requires a new
outcome cache. A-CASE's virtual basis, introduced below, instead keeps its
reference fixed while expanding its operator set.

Finally, $Z_pZ_q$ and $Y_pY_q$ commute globally but not qubit-wise.
An entangling readout can measure them together, whereas qubit-wise
readout separates them. This illustrates why grouping changes both the
setting count and the circuit cost, the trade tested in
Sec.~\ref{sec:admitted}.

\subsection{When \texorpdfstring{odd-$Y$}{odd-Y} filtering applies}
\label{sec:operator-parity}

For a Hamiltonian and density operator that are real in the computational
basis, the initial gradient of an even-$Y$ Pauli rotation vanishes.
Transposition changes a Pauli word's sign according to its number of $Y$
factors; an even-$Y$ word makes $G_P$ purely imaginary and antisymmetric,
whose trace pairing with real symmetric $\rho$ is zero. Odd-$Y$ rotors
preserve a real state. Thus an exact gradient-based ADAPT run can omit
even-$Y$ candidates while it follows this real-state trajectory and stops
when all gradients vanish \cite{utama2026operatorcentric}. Finite-shot
rankings need not follow the same trajectory.

The package exposes \texttt{odd\_y\_filter}, and its qubit-ADAPT chemistry
pool contains odd-$Y$ words derived from fermionic excitations. This is a
gradient-selection rule, not a general restriction on A-CASE basis
operators or a particle-sector guarantee. The molecular A-CASE path retains
whole conserving excitation operators: individual Pauli words from their
expansion need not preserve particle number or spin. Sparse support and
measurement demand must therefore be evaluated for the actual solver and
pool, rather than inferred from the algebra alone.

\section{Capabilities and workflow}
\label{sec:features}

Table~\ref{tab:features} groups the implemented features by the task they help a
user perform. These capabilities can be used independently: an operator-algebra
experiment does not require the subspace solvers, and importing the core package
does not load optional quantum software. The core runtime dependency is NumPy
\cite{harris2020numpy}; SciPy \cite{virtanen2020scipy} and the chemistry and
bridge packages are installed as separate extras.

\begin{table*}
  \caption{Implemented capabilities and their intended use. Entry points refer
  to modules or public objects in the package. This is a capability map, not a
  performance comparison.}
  \label{tab:features}
  \footnotesize
  \begin{ruledtabular}
  \begin{tabular}{p{0.16\textwidth}p{0.43\textwidth}p{0.34\textwidth}}
Task & Available features & Entry points and scope\\
\colrule Operator calculations & Pauli and fermionic operators; gates; density operators; Kraus channels; partial traces, entropy, and entanglement diagnostics; sparse exterior products, contractions, and blade tests in a fixed Jordan--Wigner frame & \texttt{MV}, \texttt{pauli}, \texttt{fermion}, \texttt{states}, \texttt{channels}; sparse support can grow during execution\\
Model construction & Spin chains; Hubbard and other lattice models; FCIDUMP integral import; effective-Hamiltonian JSON; orbital rotations & \texttt{models}; chemistry generation has optional dependencies\\
Circuit experiments & Parameterized Pauli rotors and named Clifford gates; versioned program serialization; gradients; OpenQASM export & \texttt{ir}; exporters and bridges support defined subsets\\
Numerical execution & Exact operator and dense statevector paths; particle/spin-sector calculations; finite-shot simulation; stabilizer execution & \texttt{backends}; sector and stabilizer paths have their own state and operation restrictions\\
Eigensolver comparisons & VQE and ADAPT-VQE; qubit-exchange and coupled-exchange pools and a shared-unitary multi-root ADAPT in exact sectors; fixed and adaptive operator-response subspaces; sampled and time-evolved determinant inputs; selected-CI controls; hybrid bases and projected observables & \texttt{algorithms}, \texttt{subspace}; a converged projected solve need not reach the full ground state\\
Convergence diagnostics & Second-moment blocks; true Ritz residuals with resolution floors; folded-spectrum roots, refined Ritz vectors, and inclusion intervals; opt-in convergence reports & \texttt{SecondMomentBank}, \texttt{FoldedSpectrum}, \texttt{subspace.convergence}; exact archived validation; separately bounded sampled moments and confidence reports\\
Measurement studies & Qubit-wise and block-commuting grouping; Clifford readout circuits; anticommuting-clique compilation of a fixed functional; shared outcome caches; allocation, covariance, and uncertainty estimates & \texttt{measurement}; interval guarantees depend on the estimator and declared assumptions; clique settings do not feed the shared cache\\
Repeated experiments & Cached model preparation; independent solves; optional reference validation; object or packed coefficients; streaming with recomputation & \texttt{prepared}, \texttt{pipeline}, \texttt{subspace}; the streaming row bound excludes other process memory\\
Interoperability & Stim, OpenFermion, pytket, PennyLane, and PyZX adapters; a fragment-solver callback for an external embedding loop & \texttt{bridges}, \texttt{subspace.fragment}; optional dependencies and conversion-specific checks; the embedding loop stays outside the package\\
  \end{tabular}
  \end{ruledtabular}
\end{table*}

\subsection{From a model to a subspace calculation}

A model contains a Hamiltonian, a reference program, and metadata such as the
particle sector. A backend evaluates the reference. A subspace calculation then
chooses operators $A_i$ and forms the virtual basis
$\ket{\phi_i}=A_i\ket{\psi}$. Its overlap and Hamiltonian matrices are
\begin{align}
 S_{ij} &= \avg{\psi|A_i^\dagger A_j|\psi}, \\
 H_{ij} &= \avg{\psi|A_i^\dagger H A_j|\psi},
 \qquad Hc=ESc.
 \label{eq:projected}
\end{align}
The basis states need not be prepared separately: all matrix elements are
expectations on the same reference. A matrix-element bank caches the operator
products and their values. Fixed constructions expose the effect of a chosen
basis; adaptive growth selects additional directions while reporting energy
changes, overlap conditioning, and new measurement support. Projected
observables use the same basis and can return expectations or transitions
without constructing a full Ritz state.

The Adaptive Clifford-Algebra Subspace Eigensolver (A-CASE) is the adaptive
operator-response solver. Its candidate families include
Pauli, commutator, Krylov, and configuration directions. VQE and ADAPT-VQE
provide circuit-based comparisons \cite{peruzzo2014vqe,grimsley2019adapt}.
A separate exact-sector path compares qubit-exchange and coupled-exchange pools
\cite{ramoa2025ceo} and a uniform-weight multi-root ADAPT with one shared
unitary \cite{grimsley2025excited}; it stores dense sector matrices and does not
route through the circuit program.
Quantum-selected configuration interaction (QSCI) and sample-based quantum
diagonalization (SQD) instead select determinants and build their projected
Hamiltonian classically \cite{kanno2026qsci,robledomoreno2025sqd}; their
samples can also come from a short real-time evolution of the reference
\cite{mikkelsen2025teqsci}. Selected-CI
controls \cite{huron1973cipsi} and hybrid bases help distinguish the benefit of
sampling from that of basis construction.
For determinant methods, zero measured matrix-element words does not mean zero
cost: state preparation, sampling yield, matrix construction, and diagonalization
still matter.

The repository examples provide entry points for exact algebra, adaptive
subspaces, finite-shot reconstruction, sector execution, and effective models.
A useful first experiment is to hold the model and reference fixed, compare
fixed and adaptive bases, then sample their matrix elements. This separates
subspace error from measurement error before introducing a device cost model.

\subsection{Residuals and folded-spectrum roots}
\label{sec:residuals}

For an unregularized Ritz solve, the residual is orthogonal to the retained
span in exact arithmetic. A small \emph{projected} residual therefore does not
establish a small full-space residual, and an adaptive stopping reason alone
does not establish eigenstate convergence. The pair $(S,H)$ must be supplemented
by a second-moment block on the same reference,
\begin{equation}
 K_{ij}=\avg{\psi|A_i^\dagger H^2 A_j|\psi},
 \label{eq:second-moment}
\end{equation}
which \texttt{SecondMomentBank} builds from the bank's projected-observable rows
for $H^2$. Squaring the projected Hamiltonian would discard excursions outside
the span and cannot supply this diagnostic. For a nonzero
$\ket{\Psi}=\sum_j c_j A_j\ket{\psi}$ and a real target
$\omega$,
\begin{align}
 \mu(\omega)&=\frac{\lVert(H-\omega)\Psi\rVert^2}{\avg{\Psi|\Psi}}
 =\frac{c^\dagger F(\omega)\,c}{c^\dagger S c}, \nonumber\\
 F(\omega)&=K-2\omega H+\omega^2 S.
 \label{eq:folded}
\end{align}
At the Rayleigh quotient $E$ this is the residual variance, $\sigma^2=\mu(E)$.
More generally, $\mu(\omega)=\sigma^2+(E-\omega)^2$: the folded objective
penalizes both variance and displacement from the target.
The computed $\mu$ and $\sigma^2$ can suffer cancellation between large terms,
so they are reported signed and count as resolved only above a fixed relative
threshold times their cancellation scales. A rounded zero stays unresolved
rather than becoming a converged residual. This threshold is a numerical
convention, not a proved floating-point error bound. Before the block is built,
a storage gate bounds its coefficient occurrences from support sizes alone;
that is an admission rule, not a byte or peak-memory bound.

The same three blocks give the projected folded-spectrum operator
\cite{wang1994folded,cadi2024folded}. The lowest root of $F(\omega)c=\mu Sc$
minimizes $\mu(\omega)$ over the retained span; it need not identify the exact
eigenstate nearest the target. Setting $\omega$ to a Ritz value gives a refined
Ritz vector, a standard residual-minimizing construction \cite{jia1997refined}.
Updating the target to the resulting Rayleigh energy gives a variance-minimizing
iteration. Each target requires one small generalized eigensolve and no new rows.
A folded solve keeps the span a Ritz solve with the same thresholds would keep,
and it refuses an overlap ridge, which would change the metric. Because $\mu$ is
a weighted mean of $(E_k-\omega)^2$ over the eigenvalues the state touches, in
exact arithmetic some
eigenvalue lies within $\sqrt{\mu}$ of $\omega$ and, at the Rayleigh quotient,
within $\sigma$ of $E$, the Weinstein interval \cite{weinstein1934modified}; the
package adds the respective resolution allowance under each square root when
reporting these intervals. This guards against cancellation under the stated
numerical convention; it is not a verified numerical enclosure. Neither interval says
\emph{which} eigenvalue it holds. A ground-state claim needs separately
established spectral input, which convergence reports accept and never infer.

Opt-in convergence reporting attaches the exact residual to each tracked root at
adaptive termination, without changing selection or stopping, and enters its
time and resident rows in the bank's resource ledger. On the \cqcResidualBanks{}
frozen banks the validation records cover, the second-moment block, every Ritz
residual and every refined folded root agree with dense statevector oracles.
Those archived validations use exact reference pairings. A separately bounded
sampled extension now reconstructs overlap, Hamiltonian and second-moment
blocks with simultaneous entrywise confidence, reports unresolved normalization
and cancellation, and supplies physical-direction folded inclusion intervals.
Its engineering validation is distinct from the historical cost programme in
Sec.~\ref{sec:residual-closure} and makes no advantage claim.

\subsection{Reusing preparation across experiments}

Repeated molecular experiments often share the same integrals and fermionic
mapping. The command-line pipeline makes that common work a
\texttt{prepare} stage, followed by independent \texttt{solve} invocations
and optional \texttt{validate} runs. Each stage is available through
\texttt{python -m clifford\_qc.pipeline} and can run in its own process.
Preparation, a large matrix-element bank, and a reference eigensolve
therefore need not coexist in memory.

\subsection{A molecular run through the pipeline}
\label{sec:molecular}

Consider the repository's minimal-basis H$_4$ fixture, a restricted real
FCIDUMP file \cite{knowles1989fcidump}. The orbital, active-space, and
integral calculation precedes this pipeline; an electronic-structure
package such as PySCF can supply it \cite{sun2018pyscf}. The committed
fixture allows the following stages to be exercised without chemistry
extras.

\begin{table*}
  \caption{Stages of the molecular command-line pipeline. Each row is a
  separate invocation. Reuse applies to the declared inputs and settings;
  the table describes interfaces, not measured performance.}
  \label{tab:pipeline}
  \footnotesize
  \begin{ruledtabular}
  \begin{tabular}{@{}p{0.10\textwidth}p{0.47\textwidth}p{0.37\textwidth}@{}}
 Stage & Reads and writes & Reuse scope \\
 \colrule
 \texttt{prepare} & Reads the integral file and preparation options; writes a content-addressed model artifact and reports its path, fingerprint, and cache-hit status & Reused when source bytes, sector overrides, tolerance, model name, encoding, implementation fingerprint, and NumPy version match \\
 \texttt{solve} & Reads the artifact and A-CASE/storage options; writes energy history, selected labels, stopping reason, resource diagnostics, elapsed time, and fingerprints & Reuses preparation; builds a separate bank and run record for each configuration \\
 \texttt{validate} & Reads the artifact and reference-solver options; writes reference energies, eigenpair residuals, sector dimension, and elapsed time & Reference results can be reused for unchanged model, sector, and reference settings; their cost remains separate \\
  \end{tabular}
  \end{ruledtabular}
\end{table*}

First, \texttt{prepare} parses the file, maps its integrals to Pauli words,
and stores the ordered Hamiltonian, reference program, and model metadata.
It runs no eigensolver. Its JSON response identifies the artifact to pass
to the next stage. Repeating the call with unchanged inputs validates the
stored digest and preparation key before skipping parsing and mapping.
Different active-space integrals produce a different artifact. Consumers
receive fresh model objects; the implementation fingerprint is memoized
within a process, so source changes require a restart.

Next, the solver evaluates a determinant reference directly, without
materializing its density multivector, and grows an A-CASE basis from whole
determinant excitations. The prepared molecular path uses Jordan--Wigner
occupation bits and returns energy, selection and resource records. The exact
residual of Sec.~\ref{sec:residuals} is an opt-in Python diagnostic; the CLI
record does not carry it.

A user can compare storage policies by repeating this prepared solve on the
same artifact, choosing \texttt{object} or \texttt{packed} storage independently
of \texttt{retain\_all} or \texttt{stream\_recompute} lifetime policy.
These choices change bank storage, not the reference-state representation.
The bank is rebuilt for each invocation; only preparation is cached across
the runs. The historical process-isolated storage study described below used
its own benchmark worker and the legacy density reference in every arm, not
the present determinant-reference default. Its measurements do not compare
the reference representations. Python callers can opt into oracle-based stopping, which is identified in the run record.

Finally, optional \texttt{validate} uses the sector statevector backend to
compute reference eigenpairs. Dense validation suits the small fixture;
matrix-free Lanczos or the optional SciPy eigensolver serves larger sectors.
Reported residuals belong to its eigenpairs, not to the A-CASE state's
full-space residual. A comparison that uses validation must account for its
separately recorded time.

The complete shell commands, input path, and output fields are documented
in \texttt{molecular/PIPELINE.md}. Finite-shot methods, other eigensolvers,
and multi-root adaptation use Python interfaces rather than this CLI.
The process-isolated storage measurements in Sec.~\ref{sec:storage} quantify a
peak-memory reduction on the larger tested inputs at increased runtime. They
establish no production molecular speedup or asymptotic scaling advantage.

\section{Architecture}
\label{sec:architecture}

The architecture separates model construction, execution, solvers, measurement,
and experiment records. A shared algebra keeps operator conventions consistent,
while explicit interfaces let these layers evolve at different rates.

\subsection{Composing calculations}

\begin{figure*}
  \includegraphics[width=\textwidth]{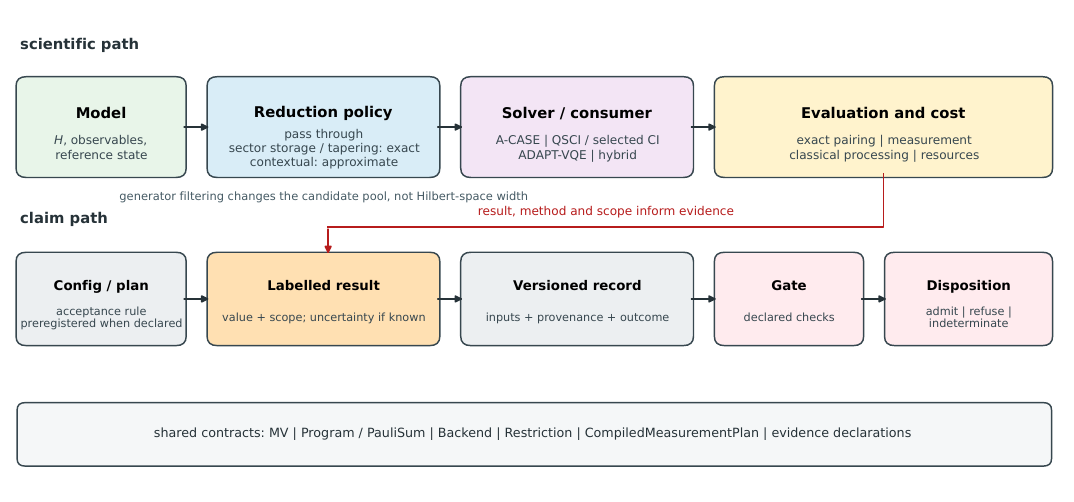}
  \caption{The end-to-end architecture, separating a scientific result from a
  publishable claim.  The scientific path constructs a model, applies only
  explicit reductions, selects a solver policy, and evaluates an exact or
  measured estimator.  The claim path attaches evidence, writes provenance,
  and lets a declared gate admit, refuse, or leave the result indeterminate.
  The band at the foot names the typed contracts shared across both paths.
  Generator filtering changes the candidate pool; it is not shown as a
  Hilbert-space reduction. This is a schematic of possible stages, not a
  mandatory execution order: sampled determinants can also be inputs to a
  solver, and not every result passes through an independent claim gate.}
  \label{fig:architecture-flow}
\end{figure*}

Figure~\ref{fig:architecture-flow} makes explicit a distinction a layer
inventory alone cannot show.  The reductions available in \pkg{} are not
interchangeable, and the architecture keeps them apart because their scopes
differ.  Sector representation compresses exact storage; symmetry tapering
reduces the active register exactly; contextual restriction reduces it
approximately and carries the removed Hamiltonian fraction as a bias source;
and A-CASE or determinant projection reduces only the subsequent generalized
eigenproblem.  Generator filtering changes which directions may enter that
projected solve, not the Hilbert space itself.  Collapsing any two of these into
one ``reduction'' step would make the resulting error budget unattributable,
which is the first place scope would be lost.

\subsection{Provider workflows and evidence transport}

The restart architecture treats the Clifford algebra as an operator language,
not a requirement to materialize a density multivector. An opt-in workflow
wraps existing exact-projection, sampled-projection and sector-action providers
in a versioned result. It does not replace adaptive solvers or make the spans
of the provider paths equivalent. Exact projection is not a ground-state or
true-residual certificate. Sampled projection supplies simultaneous entrywise
bounds under its declared assumptions, not a confidence interval for the
selected Ritz energy. Sector-action residuals establish numerical convergence
within their sector, not that the root is the ground state.

In the prepared adaptive path, the existing convergence and
reference-conditioned certification consumers are still coupled to the legacy
matrix-element bank and its density-reference interface; they have not yet been
ported to the determinant provider. This is an implementation boundary, not a
mathematical need for a density expansion. The determinant stores occupation
bits instead of the determinant projector's $2^n$ Pauli coefficients; the bank,
operator products and temporary work remain additional costs. Table~\ref{tab:storage}
measures full-process memory only for the legacy path, so it cannot quantify
RSS or runtime for the default path without a matched process-isolated study.

A scoped ledger times the wrapper's sampling and solve calls, while provider
preparation and independent validation can be supplied as caller-reported
costs. Export is separate; no end-to-end total, process peak memory or hardware
time is inferred. A portable archive carries the result, a digest manifest and
explicitly supplied attachments. Its verification checks integrity, but does
not authenticate a publisher, establish preregistration, supply missing source
or environment inputs, or independently validate scientific claims. The caller
retains an external digest and arranges persistent storage.

The shared experiment runner instead records execution plans, task outcomes,
checkpoints and immutable content-addressed artifacts for CI and replays. Its
transport receipts are not scientific cost or advantage claims. Numerical
replay compares recorded fields under the original tolerances and reports
execution-fingerprint drift separately from strict replay. These interfaces and
their limitations are specified in the repository's restart documentation.

\subsection{The kernel}

The numerical kernel's canonical operator type is the multivector \texttt{MV}:
the sparse container of Sec.~\ref{sec:operator}, a map from packed Pauli-word
codes to complex coefficients carrying the exclusive-or product described
there. Density operators, gates, individual Kraus operators, and fermionic
creation and annihilation operators can all use this type.  The public program intermediate
representation exposes a distinct \texttt{PauliSum} container for Hamiltonians
and observables; its
\texttt{to\_mv} and \texttt{from\_mv} boundary is exact and uses the same packed
word codes.  It is a real container conversion, and it is deliberately not a
change of algebraic basis or operator semantics.  Benchmarks therefore need not
own a basis-translation convention, while the architecture still names the
boundary the public interface actually has.

In the fixed Jordan--Wigner frame, the kernel also offers the exterior
structure of the same algebra on sparse operands: wedge products, left and right
contractions, an inverse of the generator-subset map that does not rebuild an
ordered blade's phase, and a blade test that checks the Pl\"ucker condition
through the active contraction masks of the stored terms rather than by
enumerating every code of the grade. These bound the work of one test. They do
not bound the support that later products reach, and they are classical kernel
operations, not a quantum resource saving; a separate fixture audit rebuilds
their decisions and records timings without speed thresholds.

For homogeneous elements of grades $r$ and $s$, the exterior product is
$A_r\wedge B_s=\langle A_rB_s\rangle_{r+s}$ in the fixed Majorana frame.
For a nonzero homogeneous positive-grade $k$-vector $A_k$, the full Pl\"ucker simplicity
condition tests
\begin{equation}
  (\beta\mathbin{\lrcorner} A_k)\wedge A_k=0.
  \label{eq:plucker}
\end{equation}
The condition must hold for every basis blade $\beta$ of grade immediately below
$k$. The left contraction is the grade-selected product with reversion, as
defined in the conventions. The implementation tests only masks formed by
deleting a generator from an occupied support mask; all other basis
contractions vanish. Its tolerance makes this a numerical test. Subset masks
identify generators, while the ordered product retains its Pauli phase.

The reference-state pairing in Eq.~\eqref{eq:projected} serves a different
purpose from the Hilbert--Schmidt pairing used to measure removed Hamiltonian
norm. A small projected energy error does not follow from a small operator norm
fraction alone. The restriction and validation records keep these quantities
separate.

\subsection{Package layers}

\begin{figure*}
  \includegraphics[width=\textwidth]{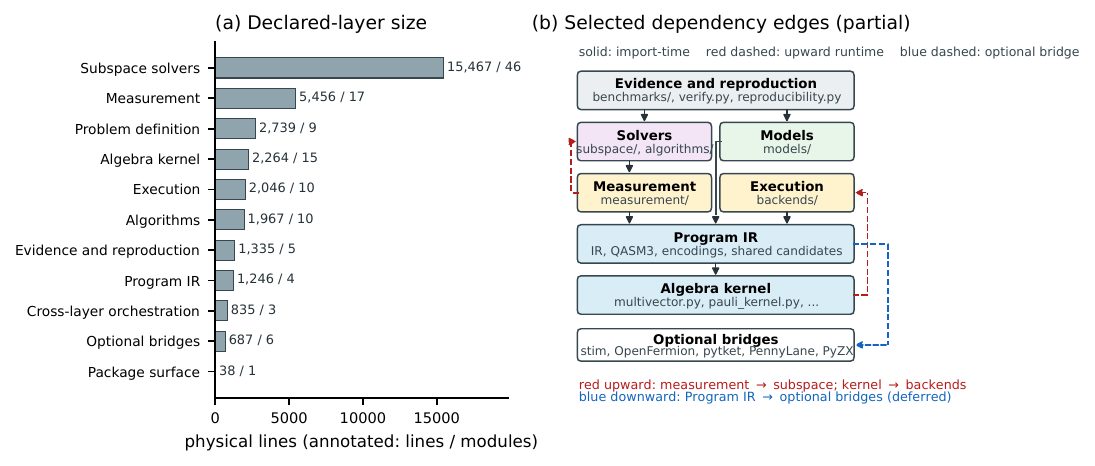}
  \caption{The package as measured.  (a) Physical lines per declared layer,
  annotated with lines and module count; the census is recomputed from the
  source tree whenever this manuscript is rebuilt.  (b) A partial dependency
  view with shortened layer labels; the solver box groups algorithms and
  subspace solvers, while the package surface and cross-layer
  orchestration are omitted from this selected-edge view.  Solid arrows are
  selected import-time dependencies, including the downward chemistry-to-IR
  dependency for shared named Pauli candidates. Red dashed arrows are genuine upward runtime edges deferred to
  call time; the blue dashed arrow is a downward call into an optional bridge.
  The deferral buys initialization order, not independence, and
  Sec.~\ref{sec:architecture} states the exceptions rather than claiming an
  acyclic runtime graph.}
  \label{fig:layers}
\end{figure*}

\begin{table}
  \caption{Declared layers of \pkg{}, recomputed from the source tree at
  manuscript build time.  The share column is of package lines; the test suite
  is listed for scale and is not a layer.}
  \label{tab:layers}
  \begin{ruledtabular}
  \begin{tabular}{lrrr}
  Layer & Modules & Lines & Share \\
  \colrule
  \input{tables/layers.tex}
  \end{tabular}
  \end{ruledtabular}
\end{table}

Figure~\ref{fig:layers} and Table~\ref{tab:layers} summarize the current
implementation: \cqcModules{} modules and \cqcLines{} physical lines, with
\cqcTestModules{} test modules and \cqcTestLines{} test lines. The kernel
contains \cqcKernelLines{} lines, the subspace layer \cqcSubspaceLines{}, and
the measurement layer \cqcMeasurementLines{}. These counts describe the code's
distribution; they do not measure correctness or software quality.

The layer map is maintained explicitly. The table generator checks that every
package module belongs to exactly one declared layer and fails on an unassigned
module. This keeps the inventory current, but it does not prove that all imports
obey the diagram.

\subsection{Interfaces and dependencies}

The main interfaces are \texttt{MV}, the program intermediate representation,
backend protocols, grouped samples, models, labelled basis directions, compiled
measurement plans, and symmetry restrictions. The program representation uses
Pauli rotors $\exp(-\mathrm{i}\theta P/2)$ and named Cliffords. The grouped-sample
contract represents both qubit-wise settings and Clifford-diagonalized settings
with explicit keys and signed readout maps, allowing both forms of measurement
to feed shared estimators. Sector execution uses a specialized statevector API;
it is not a drop-in replacement for every density-operator path.

Dependencies are mostly layered, with explicit exceptions. Chemistry model
construction and algorithm pools share the candidate type and odd-$Y$ predicate
in \texttt{pool\_operator}, assigned to Program IR because it depends on
\texttt{PauliWord}. The historical algorithm imports re-export the same type;
model construction no longer imports the solver package for those values. Some measurement
methods call subspace code, and kernel sector helpers call execution code,
through function-scope imports. These deferred imports manage initialization;
they do not make the runtime dependency graph acyclic. Optional adapters are
imported by name rather than loaded by the package initializer. There are
\cqcBridgeModules{} bridge modules, including their package initializer.

\section{Uncertainty, device cost, and memory}
\label{sec:scopes}

\begin{figure*}
  \includegraphics[width=\textwidth]{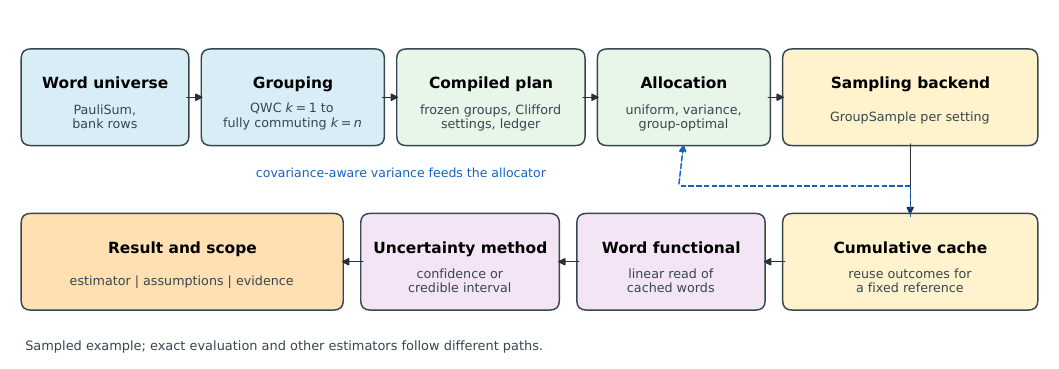}
  \caption{A sampled measurement workflow, from a word universe to an
  uncertainty statement. Shared outcomes are reusable only for compatible
  readout and the same reference; additional shots still incur cost. Exact
  pairings and other estimators can follow different paths. Evidence depends on
  the method and assumptions and is not established by a label alone.}
  \label{fig:evidence}
\end{figure*}

Three kinds of information accompany numerical results: how they were estimated,
which device assumptions a cost uses, and which allocations a memory report
covers. The following interfaces preserve this context, with different levels of
enforcement.

\subsection{Shared measurements and uncertainty}

Figure~\ref{fig:evidence} shows one pass.  A word universe is grouped, either
qubit-wise or under dyadic block commutativity; the grouping is frozen into a
compiled plan with signed parity readouts, a matched resource ledger, and
Clifford settings synthesized through a stabilizer tableau
\cite{gidney2021stim,verteletskyi2020grouping,crawford2021efficient,miller2024hardware};
shots are allocated; the sampling backend returns joint outcome histograms per
setting; a cumulative cache reuses outcomes across observables on the same reference; a
linear functional reads the cache; and a confidence bound turns the read into an
interval with a stated family and failure probability
\cite{maurer2009empirical,sidak1967rectangular}.

The main evidence categories are exact, asymptotic, finite-sample, heuristic,
and none. Here ``exact'' means a calculation without shot noise, within its
stated numerical and model assumptions. It does not certify that a selected
subspace contains the full ground state. A finite-sample bound applies to its
specified functional and sampling conditions; it is not automatically a bound
on the nonlinear Ritz eigenvalue.

Subspace bias and sampling uncertainty can contribute to the same accuracy
budget, but their origins and guarantees must remain explicit. Cost comparisons
therefore retain evidence categories and distinguish an oracle comparison from
a statistical certificate. Some result labels are conventions rather than
constructor-validated values, as the record census in Appendix~\ref{app:evidence} documents.

The versioned evidence contract records a tier and scope per quantity;
it does not itself verify the scientific value. The record census and treatment
of historical annotations are given in Appendix~\ref{app:evidence}.

\subsection{Device cost and admissibility}

Table~\ref{tab:device-cards} lists the scenario inputs used by the committed
cost study. For a setting with one- and two-qubit gate counts $N_a,N_b$ and depths
$D_a,D_b$, the model uses routed values $N'_b=\lceil m_N N_b\rceil$ and
$D'_b=\lceil m_D D_b\rceil$. With $n$ measured qubits, it assigns
\begin{align}
 t_g &= t_{\rm prep}+t_aD_a+t_bD'_b+t_{\rm read}+t_{\rm reset},\\
 F_g &= (1-\epsilon_a)^{N_a}(1-\epsilon_b)^{N'_b}
       (1-\epsilon_{\rm read})^n.
 \label{eq:device-model}
\end{align}
The setting is admissible when $F_g$ meets the floor. The multipliers are
surrogates applied to counts and depths, not a routed circuit or a connectivity
graph; logical preparation-circuit gates beyond the stated preparation time
are outside the setting ledger. Raw-shot schedules are priced as
$\sum_g S_g t_g$; where used, the effective-shot surrogate is $S_gF_g^2$.

\begin{table*}
  \caption{Illustrative device-card inputs from the committed cost record.
  Times are in microseconds; errors are independent per-operation surrogates.
  The first card is a logical baseline, the others design scenarios. All-to-all
  applies to the first and second rows; the third uses a sparse-connectivity
  surrogate, not an explicit graph. $m_N,m_D$ scale logical two-qubit count and
  depth respectively before ceiling.}
  \label{tab:device-cards}
  \scriptsize
  \compactcolumns
  \begin{ruledtabular}
  \begin{tabular}{lrrrrrrrrrrr}
  Card & $t_{\rm prep}$ & $t_a$ & $t_b$ & $t_{\rm read}$ & $t_{\rm reset}$ & $\epsilon_a$ & $\epsilon_b$ & $\epsilon_{\rm read}$ & $m_N$ & $m_D$ & Floor \\
  \colrule
  \input{tables/device_cards.tex}
  \end{tabular}
  \end{ruledtabular}
\end{table*}

Device cards specify connectivity, gate times, a fidelity floor, and calibration
status. The cost path evaluates already-synthesized measurement settings under
these assumptions. If a required setting falls below the fidelity floor after
applying the routing surrogate, the ledger marks the protocol
\emph{inadmissible} and leaves its runtime absent. Downstream comparisons must
respect that status.

The module prices settings; it does not choose a subspace or allocate shots to
an accuracy target. Its independent-error fidelity model and scalar routing
surrogate are accounting assumptions, not a hardware calibration. The
non-baseline cards explicitly identify themselves as illustrative project
scenarios. Section~\ref{sec:admitted} shows how these assumptions change the
interpretation of a measured shot reduction.

\subsection{Coefficient storage and recomputation}
\label{sec:storage}

Matrix-element banks offer object and packed coefficient representations.
Packing changes storage, not the operator basis. Direct support queries and
exact contractions avoid reconstructing multivectors merely to score a
candidate or report storage. Measurement functionals can own packed snapshots
that remain usable after bank rows are evicted.

The default policy retains every constructed pair. The opt-in streaming policy
retains the selected block and a bounded frontier of other pairs, rebuilding
rows when needed. Persistent overlap and Hamiltonian coefficient rows are
bounded by twice the retained-block pair count plus twice the frontier limit.
Two additional product rows can exist temporarily during construction. Packed
streaming rows own independent buffers, so eviction releases their capacity.

This is not a total-process memory bound. Support histories, interned words,
scalar caches, generator products, the reference state, and explicitly requested
observables have other lifetimes. Resource output reports resident and cumulative
quantities separately; its component diagnostics are not an additive estimate
of process resident memory. Streaming can reduce resident coefficient storage at
the cost of recomputation. Neither packing nor streaming is a universal speed
improvement, and the retain-all object path remains the default.

\begin{table*}
  \caption{The coefficient-storage arms measured in isolation. Each cell is
  one run of the same adaptive solve in a fresh interpreter, with full
  candidate pools, a common basis budget and the same stopping rule; every arm
  uses the legacy density reference. $M$ is the final basis size, identical
  across the arms. Peak memory is the worker's
  own high-water resident set, including preparation, support histories and
  instrumentation; wall time is the supervising parent's. Single observations
  on one host, not a memory bound.}
  \label{tab:storage}
  \footnotesize
  \begin{ruledtabular}
  \begin{tabular}{lrrrrrrrrr}
  & & \multicolumn{4}{c}{Peak process memory (GiB)} & \multicolumn{4}{c}{Wall time (min)} \\
  & & Object & Packed & Object & Packed & Object & Packed & Object & Packed \\
  System & $M$ & retain & retain & stream & stream & retain & retain & stream & stream \\
  \colrule
  \input{tables/storage_policy.tex}
  \end{tabular}
  \end{ruledtabular}
\end{table*}

Table~\ref{tab:storage} measures these choices on the historical density-MV
path, not on the determinant-reference default. Each of the
\cqcPolicyArms{} arms runs on \cqcPolicySystems{} fixed molecular inputs in its
own process, so no arm inherits another's allocator state. Every arm reproduces
the same selected labels, energy histories and stopping reasons, and an independent
checker rebuilds the final projected pencils by direct products. Leaving out
H$_4$, whose observed peaks differ by at most \cqcHFourPolicySpread{}, packing alone lowers
the observed peak process memory by \cqcPackSaveMin{}--\cqcPackSaveMax{},
streaming object rows by \cqcStreamSaveMin{}--\cqcStreamSaveMax{}, and both
together by \cqcBothSaveMin{}--\cqcBothSaveMax{}. These are separate measured
outcomes, not factors that multiply. The default object-retain arm is the
fastest on every system: packing and streaming buy memory with time. The peaks
are process observations, not an attribution to coefficient rows; the small
separation on H$_4$ likewise does not identify which component dominates its
memory. Every arm
completed under the memory ceiling the campaign declared, so on these inputs
neither option is needed for feasibility. The matrix establishes a measured
memory--time trade on these \cqcPolicySystems{} inputs; it establishes no
universal memory bound and no production speedup, and it leaves the object
default unchanged.

\section{Producers, gates, and preregistration}
\label{sec:gates}

Benchmark producers write observations; independent checks may recompute values,
verify preregistered configurations, or audit provenance and structural
consistency. Preregistration fixes a decision rule before new observations;
post-hoc declarations retain their different evidential status. Not every record
has an independent numerical replay. Appendix~\ref{app:gates} describes the
validation coverage and preserves the full gate census.

\begin{table*}
  \caption{The dyadic block-commuting hierarchy on two committed banks. The
  last column is the break-even threshold for the cost per logical CX divided
  by the cost per preparation: settings saved divided by added logical CX
  applications per sweep, relative to $k=1$. It is undefined at $k=1$.
  All-to-all logical Clifford circuits, no routing, noise or mitigation.}
  \label{tab:hierarchy}
  \begin{ruledtabular}
  \begin{tabular}{llrrrrr}
  Bank & $k$ & Settings & Mean CX per setting & Max CX depth & Preparations at uniform shots & Break-even $c_{\rm CX}/c_{\rm prep}$ \\
  \colrule
  \input{tables/hierarchy.tex}
  \end{tabular}
  \end{ruledtabular}
\end{table*}

\begin{figure*}
  \includegraphics[width=\textwidth]{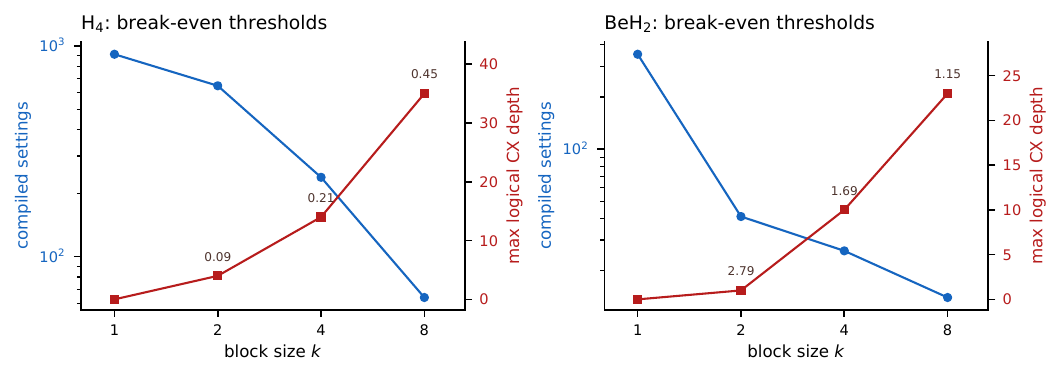}
  \caption{Compiled settings (left axis, blue) and maximum logical entangling
  depth per setting (right axis, red) along the block-size axis, for the two
  committed banks. Annotations give the break-even threshold for cost per CX
  divided by cost per preparation, relative to $k=1$. A rung saves cost under
  the stated model only when the actual unit-cost ratio is below its threshold.
  Maximum depth is distinct from total CX count.}
  \label{fig:hierarchy}
\end{figure*}

\begin{table*}
  \caption{Qubit-wise against fully commuting measurement on one frozen
  BeH$_2$ bank, from the preregistered comparison.  The ratio column is the
  fully commuting value over the qubit-wise value.  Runtimes are
  accuracy-matched logical accounting under illustrative device cards, not
  hardware measurements.}
  \label{tab:phase14b}
  \begin{ruledtabular}
  \begin{tabular}{lrrr}
  Quantity & QWC ($k=1$) & Full ($k=8$) & Ratio \\
  \colrule
  \input{tables/phase14b.tex}
  \end{tabular}
  \end{ruledtabular}
\end{table*}

Reproduction also depends on the numerical environment.  Continuous integration pins one thread for the
numerical libraries, because threaded basic-linear-algebra reductions sum in a
thread-count-dependent order and the same symmetric eigensolve returns different
last bits across thread counts.  This controls one source of numerical variation.  It does not make records portable across processor
microarchitectures, because kernel selection by instruction set remains, and the
reproduction documentation records that floor rather than claiming bit-exact
portability \cite{peng2011reproducible,wilson2017practices}.

\section{Measurement and cost results}
\label{sec:admitted}

The following studies exercise grouping, shot allocation, and device cost on
small molecular banks. They show why a setting count, a shot count, and a runtime
answer different questions. For orientation, QWC denotes qubit-wise commuting
measurement, full commutation allows entangling readout circuits, CX denotes a
controlled-NOT gate, and mHa denotes millihartree. A rung is one allowed block
size in the grouping hierarchy.

\subsection{Grouping trades settings for circuit depth}

Commuting-group measurement is usually presented as a setting-count reduction.
Under the dyadic hierarchy --- two words are compatible when their restrictions
commute inside every contiguous $k$-qubit block, so $k=1$ is exactly qubit-wise
commutation and $k=n$ is full commutation --- the reduction is real and large.
On the H$_4$ bank, compiled settings fall from \cqcHFourSettingsQWC{} to
\cqcHFourSettingsFull{}, a preparation reduction of \cqcHFourPrepReduction{}.
On the BeH$_2$ bank they fall from \cqcBeHSettingsQWC{} to
\cqcBeHSettingsFull{}, a reduction of \cqcBeHPrepReduction{}.

Table~\ref{tab:hierarchy} and Fig.~\ref{fig:hierarchy} also report circuit
cost. Under the simple model $C_k=R(c_{\rm prep}G_k+c_{\rm CX}N_k)$,
where $G_k$ is settings per sweep, $N_k$ is logical CX applications per sweep,
and $R$ is uniform shots per setting, rung $k$ is cheaper than qubit-wise
readout only when $c_{\rm CX}/c_{\rm prep} < (G_{\rm QWC}-G_k)/N_k$. The right-hand
side is a break-even threshold, not an observed cost ratio. The model omits
routing, noise, and other costs.

For H$_4$ the full-commuting threshold is \cqcHFourCXRatio{}, corresponding to
\cqcHFourCXAvoided{} logical CX applications per avoided preparation. For
BeH$_2$ it is \cqcBeHCXRatio{} at full commutation and
\cqcBeHCXRatioWorst{} at $k=2$. Thus the second bank tolerates a higher CX-to-
preparation unit-cost ratio in these comparisons. No saving follows without
specifying that ratio; setting reduction alone is insufficient.

\subsection{Shot reduction for a fixed energy functional}

\begin{table*}
  \caption{Anticommuting cliques against qubit-wise grouping on fixed
  Hartree--Fock reference energies, from the preregistered comparison.
  Two-qubit gates are summed over the compiled settings.  The certified
  endpoint is the first point of the fixed effective-shot grid whose total
  radius meets the target, and $>$ marks an arm censored at the top of the
  grid.  $C_{\mathrm{time}}(\epsilon)$ prices that endpoint under each
  illustrative card; a censored arm has no price, and an inadmissible one has no
  runtime. Prices cover the endpoint schedule, excluding other grid looks,
  covariance audits, and classical compilation. The decision concerns effective
  shots, per instance; nothing is
  pooled.}
  \label{tab:phase19}
  \footnotesize
  \tightcolumns
  \begin{ruledtabular}
  \begin{tabular}{llrrlrrrp{0.17\textwidth}}
  & & & Two-qubit & Certified & \multicolumn{3}{c}{$C_{\mathrm{time}}(\epsilon)$ (s)} & \\
  System & Arm & Settings & gates & endpoint & logical-alltoall & ion-like & superconducting-like & Decision \\
  \colrule
  \input{tables/phase19.tex}
  \end{tabular}
  \end{ruledtabular}
\end{table*}

The preregistered comparison in
Table~\ref{tab:phase14b} fixes one BeH$_2$ bank of \cqcBankWords{} measured
words supporting a \cqcBankBasis{}-vector subspace, and asks for the total
physical shots at which the first-order Ritz energy functional is certified
inside a stochastic radius, under a two-sided empirical Bernstein bound with a
declared family and failure probability, with the same allocation rule on both
arms.

The fully commuting arm is certified at \cqcFullCertifiedShots{} total physical
shots against \cqcQWCCertifiedShots{} for the qubit-wise arm: a ratio of
\cqcShotRatio{}, or one part in \cqcShotRatioInverse{}. The Ritz functional has measured support on
\cqcRitzSupport{} of the \cqcBankWords{} assigned words. Its support and
within-setting covariance determine the allocation; the bank's total setting
count alone does not predict the shot ratio. Table~\ref{tab:phase14b}
distinguishes settings compiled for the whole bank from settings the functional
actually touches.

Then the device cards are applied, and the result splits.  On the all-to-all
logical baseline the fully commuting arm costs \cqcLogicalRuntimeRatio{} of the
qubit-wise runtime.  On the slow-gate all-to-all card it costs
\cqcIonRuntimeRatio{}: still cheaper, but most of the shot saving is consumed by
two-qubit gate time, since that arm's settings carry entangling gates and the
qubit-wise arm's carry none.  On the sparse-connectivity card the comparison
does not happen at all.  \cqcSCInadmissibleSettings{} of the fully commuting
arm's \cqcSCSettings{} compiled settings fall below the card's fidelity floor
once the routing surrogate is applied --- mean setting fidelity
\cqcSCMeanFidelity{} against a floor of \cqcSCFidelityFloor{} --- so the ledger
returns \emph{inadmissible} rather than a number.

The finite-sample result concerns the fixed first-order functional under the
recorded sampling assumptions. It is not a certificate of total ground-state
error or a demonstration on hardware. The device ledger adds a separate test of
whether the assumed circuit fidelity is acceptable.

\subsection{Anticommuting cliques for fixed reference energies}
\label{sec:cliques}

\begin{table*}
  \caption{Accuracy-matched cost of the priced BeH$_2$ instance on the
  Jordan--Wigner arm, at the \cqcTargetMilliHartree{}~mHa target.  $k^*$ is the
  set of rungs whose cost interval reaches the smallest upper bound; the status
  column is the record's own resolution state.  The spread column is over all
  five mapping arms at the same rung.}
  \label{tab:cost}
  \footnotesize
  \begin{ruledtabular}
  \begin{tabular}{llrllrlr}
  Device card & Estimator & $k^*$ point & $k^*$ region & Status & $\Ctime$ at $k^*$ (s) & Cheapest mapping & Mapping spread \\
  \colrule
  \input{tables/cost.tex}
  \end{tabular}
  \end{ruledtabular}
\end{table*}

\begin{figure*}
  \includegraphics[width=\textwidth]{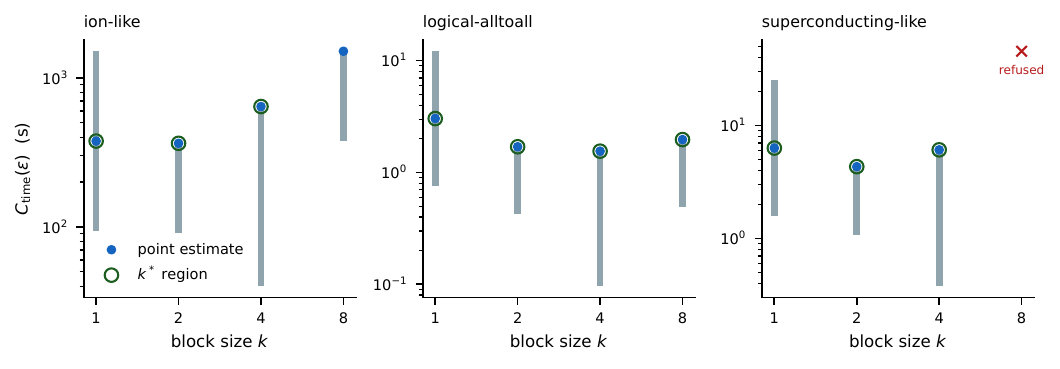}
  \caption{Accuracy-matched cost brackets for the priced BeH$_2$ instance,
  Jordan--Wigner arm, single-assignment estimator.  Grey bars span the
  confirmed-failing to confirmed-passing interval; the open circle marks
  membership in the $k^*$ region.  Each panel has its own vertical scale: the
  three cards differ by orders of magnitude in gate time, so a setting count cannot substitute for the cost model.  A rung the card refuses
  carries no cost at all and is marked at the top of its panel.}
  \label{fig:cost}
\end{figure*}

Qubit-wise and fully commuting grouping both partition a functional into
commuting sets.  A third partition reads a set of mutually
\emph{anticommuting} words in one setting: a real unit combination of such
words is itself a Pauli-like involution, which a short sequence of Pauli
rotations maps onto a single word \cite{izmaylov2020unitary}.  The package
compiles such a cover by a fixed deterministic rule and counts the rotations and
readout changes from the emitted circuit.  A clique setting returns only its
weighted sum, never the individual word means, so it cannot feed the shared
cache of Sec.~\ref{sec:scopes}: its estimand is the fixed functional and nothing
else.

The preregistered comparison in Table~\ref{tab:phase19} measures the energy
of a fixed Hartree--Fock determinant for each of \cqcPNineteenSystems{}
molecular Hamiltonians on eight qubits. Both arms share one
outcome-independent allocation rule, \cqcPNineteenLooks{} looks on a fixed
effective-shot grid, one empirical Bernstein family spanning every system, arm
and look at a total failure probability of \cqcPNineteenDelta{}, and a
\cqcPNineteenTarget{}~mHa target. Each look uses an independent schedule; the
simultaneous bound permits reporting the first passing grid point without
assuming an anytime guarantee. Effective shots count ideal measurement
outcomes; device prices separately account for the cards' fidelity inflation.
Outcomes are sampled classically from the exact reference states' measurement
distributions, not from hardware. Because the reference energies are themselves
classically available, this tests measurement protocols rather than a need for
quantum computation to obtain those energies.

The declared material-reduction rule requires at least a
\cqcPNineteenReductionFactor{}-fold reduction in the certified grid endpoint.
For a censored QWC baseline, the clique endpoint must be no greater than the
grid ceiling divided by that factor. This is a fixed-grid comparison, not a
continuous optimum or a statement about the total cost of running the campaign.

The direction depends on the instance.  On H$_4$ the clique arm, with fewer
settings, certifies inside the grid where the qubit-wise arm does not; on LiH it
has more settings and needs more shots; on BeH$_2$ it has many more settings and
still needs fewer shots.  A setting count predicts neither direction.  The cards
then split the comparison again.  The clique rotations carry entangling gates
that qubit-wise settings do not, so on the slow-gate ion-like card the clique
arm costs more device time than the qubit-wise arm wherever both are priced,
including BeH$_2$, where it needs fewer shots; and the sparse-connectivity
superconducting-like card refuses it on H$_4$ and LiH.  Each decision applies
to its own instance, and the record names no winner across cards.

\subsection{Cost at a matched accuracy target}

The deeper cost question is not which protocol uses fewer settings but which
reaches a stated accuracy for the least device time.  Writing $\Ctime$ for the
accuracy-matched logical runtime at target $\epsilon$, the search prices each
(mapping, block size) cell by bracketing the shot budget at which the replica
root-mean-square error of the selected-rank Ritz energy clears the target, with
a zero-failure gate and a one-sided bootstrap bound \cite{efron1979bootstrap}.

Table~\ref{tab:cost} and Fig.~\ref{fig:cost} report the priced BeH$_2$ instance
along the Jordan--Wigner arm, with three practical observations.  First,
the optimal rung moves with the card and the estimator: it is $k=\cqcKStarIon{}$
on the slow-gate card under single-assignment estimation and
$k=\cqcKStarLogical{}$ on the logical baseline.  The optimum depends on the cost model as well as the Hamiltonian.  Every cell resolves to a
region rather than a point --- the slow-gate cell to $k \in
\{\cqcKStarIonRegion\}$ --- and the record reports that status rather than
collapsing it to an argmin.  Second, the mapping axis is not small: at the
qubit-wise rung the spread between cheapest and dearest fermionic encoding is
\cqcMappingSpreadKOne{} on this one instance, and Table~\ref{tab:cost} reports
it again at each card's own $k^*$.  Third, the sparse-connectivity card refuses
the full-commutation rung here too, and a refused rung is absent from the
bracket rather than assigned an infinite cost.

The search is itself labelled.  The record declares its search uncertainty
heuristic --- it is Monte Carlo bracketing on a fixed shot grid, not a
certificate --- and separately declares the benchmark comparison exact-tier
against an oracle reference.  The two labels sit in one file and are not merged.
A cross-check against the independently produced shot-search record agrees on
\cqcCrossAgreeing{} of \cqcCrossChecked{} compared crossings, and the record
classifies each disagreement as environment-marginal rather than substantive.

\begin{table*}
  \caption{Mapping spread against instance spread, on every tier that measured
  it. For a positive metric, a mapping spread is the maximum over encodings
  divided by the minimum at fixed instance; an instance spread is the maximum
  over instances divided by the minimum at fixed encoding. Structural rows
  compare the largest mapping spread with the smallest instance spread. The
  accuracy rows instead use the stated point estimates or conservative supports
  on their selected cells; rows are not interchangeable.}
  \label{tab:qr3}
  \footnotesize
  \begin{ruledtabular}
  \begin{tabular}{llp{0.20\textwidth}rrp{0.22\textwidth}}
  Readout & Tier & Aggregation & Mapping & Instance & Verdict \\
  \colrule
  \input{tables/qr3.tex}
  \end{tabular}
  \end{ruledtabular}
\end{table*}

\begin{table*}
  \caption{The contextual-restriction admission screen on one BeH$_2$ bank.
  Rows are restriction rungs indexed by the number of fixed qubits.  Bias and
  word universe are given for the two contextually restricted arms; the two
  unrestricted arms are rung-independent and are stated in the last line.  The
  last column counts how many of the four arms are admissible, out of four.}
  \label{tab:r4a}
  \footnotesize
  \begin{ruledtabular}
  \begin{tabular}{rrrrrrrr}
  Fixed & Active & CS-QSE bias (mHa) & CS-QSE words & CS-A-CASE bias (mHa) & CS-A-CASE words & $H$ removed & Arms admitted \\
  \colrule
  \input{tables/r4a.tex}
  \end{tabular}
  \end{ruledtabular}
\end{table*}

\section{Negative and indeterminate results}
\label{sec:refused}

These studies identify conditions under which a proposed comparison cannot
support a conclusion. A rejected or indeterminate comparison is useful when its
acceptance rule and inputs are preserved: it shows which assumption failed and
which conclusions remain open.

\subsection{Subspace bias can make an accuracy target unreachable}

The accuracy-matched search of the previous section covers two systems, and only
one of them is priced.  The H$_4$ bank carries an exact subspace bias of
\cqcHFourBias{}~mHa against a \cqcTargetMilliHartree{}~mHa target, so no shot
budget can bring its error inside the target: the deterministic part of the
error is already outside.  The record's status for that system is not a large
cost or a censored interval but \emph{bias floor exceeds target}, and no cost is
reported for it anywhere.  The priced BeH$_2$ bank clears the same floor with
room to spare, at \cqcBeHBias{}~mHa, which is why it has a price at all.

Increasing the shot budget reduces sampling noise but does not remove the
fixed bank's bias. The cost search checks this bias floor before spending effort
on shot allocation. This conclusion comes from the separate bias calculation
and the admission rule, not from the evidence label alone.

\subsection{Mapping spread against instance spread: negative, then
indeterminate}

A natural design question for a framework supporting several fermionic
encodings \cite{jordan1928,bravyi2002fermionic,seeley2012bravyikitaev} is
whether the encoding choice matters more than the problem instance.  If it
does, an encoding-selection layer is worth building.  The question was asked on
three tiers and Table~\ref{tab:qr3} reports all of them.

Structurally, on matched greedy qubit-wise settings across the instances sharing
one grouping protocol, the widest mapping spread within a system is
\cqcStructuralMapping{} while the narrowest instance spread across systems is
\cqcStructuralInstance{}; their ratio (smallest instance spread divided by
largest mapping spread) is \cqcStructuralMargin{}. The recorded verdict is that mapping spread is not smaller than instance spread.
The same comparison on algorithm-independent word weight returns the same
verdict.  This rejects the tested hypothesis that mapping spread is smaller. It does
not establish that an automatic encoding selector improves accuracy-matched cost.

Accuracy-matched, a second instance was priced specifically to make the
comparison possible: \cqcRThreeCCells{} cells evaluated, \cqcRThreeCFinite{}
with a finite cost interval, \cqcRThreeCCensored{} right-censored at the grid
ceiling, over \cqcRThreeCSolves{} confirmatory solves with
\cqcRThreeCFailures{} failures, of which the sparse-connectivity card admits
only \cqcRThreeCSCPriced{}.  A censored cell is recorded as censored rather than
assigned an infinite cost or quietly dropped, and it licenses no wider grid.  On
that evidence the point estimates favour the mapping axis, but the conservative
supports of the same cells overlap, and the recorded verdict is
\emph{indeterminate at this shot grid}.  A targeted refinement then added
\cqcRThreeDCells{} within-grid endpoint cells chosen before any new draw, and the
supports still overlap: minimum possible mapping spread
\cqcRThreeDMappingSupport{} against maximum possible instance spread
\cqcRThreeDInstanceSupport{}, classified \emph{indeterminate after refinement}.

The refinement targeted the parent record's point-support cells rather than
all cells entering the global extrema. Its overlapping supports establish
neither ordering. Both records therefore remain indeterminate; point estimates
alone do not justify an encoding-selection rule.

\subsection{Screening approximate restrictions before sampling}

Contextual subspace methods restrict a Hamiltonian onto a stabilizer sector
chosen for the correlation it retains rather than for an exact symmetry
\cite{kirby2021contextual}, and can be tested as a preconditioner for an
operator-response subspace solver. This restriction is approximate, whereas symmetry tapering \cite{bravyi2017tapering} is exact.  The two operations therefore use separate policies in
Fig.~\ref{fig:architecture-flow}.  The
screen in Table~\ref{tab:r4a} asks whether any restriction rung admits all four
comparison arms at once under gates fixed in advance --- an admissible bias
budget of \cqcRFourAAdmissibleBias{}~mHa, a word-universe ceiling of
\cqcRFourACeiling{}, and a fixed grouping rule.

None of the \cqcRFourARungs{} rungs admits all four arms; \cqcRFourAPassing{}
rungs pass.  The failure is structural and comes from both ends.  The
unrestricted quantum-subspace-expansion reference arm overruns the
word-universe ceiling several times over, at \cqcRFourAFullWords{} words.  The
two contextually restricted arms come in far under the ceiling, some at a
handful of words.  At the first listed rung they remove
\cqcRFourARemovedFraction{} of the Hamiltonian's non-identity Hilbert--Schmidt
norm; there the CS-QSE arm carries a bias of \cqcRFourACSBias{}~mHa and the
CS-A-CASE arm \cqcRFourACSACaseBias{}~mHa, both an order of magnitude past the
admissible budget.  The unrestricted operator-response arm is the only one
admissible at every rung, at \cqcRFourAACaseBias{}~mHa over
\cqcRFourAACaseWords{} words, and one admissible arm does not make a rung: the
screen compares four arms or it compares nothing.  There is no rung where the
cheap arms are accurate enough and the accurate arm is cheap enough.

The record reports \emph{structural screen complete, no samples}. The
comparison stopped after deterministic admission checks, avoiding a sampled
campaign whose required arms already failed the stated conditions. This is a
useful way to control experimental cost without weakening the comparison.

\subsection{Residual measurement exceeds the tested cost threshold}
\label{sec:residual-closure}

Section~\ref{sec:residuals} ships the residual as an exact pairing only.
Whether a \emph{measured} second-moment block would be affordable was asked as a
cost ratio: under uniform single-assignment allocation, the shots that estimate
the ground Ritz root's residual norm over the shots that estimate its energy, at
one matched standard error, to first order in the delta method. This
linearization requires the uncertainty in $\sigma^2$ to be small relative to
$\sigma^2$; it does not resolve a near-zero residual at arbitrary precision.
A ratio within
a factor of \cqcResidualBar{} would make the residual a by-product of measuring
the energy well; beyond it, the residual is a measurement of its own.  Both
designs that asked were declared post hoc --- the first after an earlier attempt
at the same question had been inspected and recorded as invalid, the second
after the first had reported --- and their records say so.

Under qubit-wise grouping the ratio runs from \cqcResidualQWCMin{} to
\cqcResidualQWCMax{} on the \cqcResidualQWCBanks{} banks that carry an exact
second-moment block.  Fully commuting grouping lowers every ratio on the
\cqcResidualFCBanks{} banks of eight qubits it was asked of, to
\cqcResidualFCMin{}--\cqcResidualFCMax{} under the shipped greedy cover, while
an alternative cover reads \cqcResidualFCAltMin{} to \cqcResidualFCAltMax{};
none reaches the bar.  Both verdicts read \emph{none}: the uniform
single-assignment measured block closes negative at both grouping endpoints the
package ships, on these banks, and so the measurement layer carries no
second-moment path. The threshold is the campaign's decision rule, not a
physical impossibility bound. The ratios are first-order costs computed from exact
reference variances rather than sampled data.  They say nothing about other
allocations, estimators or banks, each of which would need its own declaration.

\subsection{Pool filtering can duplicate an existing construction}

A geometric-algebra structural preconditioner was proposed as a pre-encoding
filter chain.  Its falsifier was declared before the measurement: if the filters
reproduce the existing post-encoding filters and the action-equivalence quotient
removes nothing, the proposal is a reformulation and the work scoped behind it
does not proceed.

The outcome depends on the candidate pool.  The quotient removes \cqcGOneMajorana{}
candidates from the wide Majorana-monomial pool and \cqcGOneExcitation{} from
the determinant-excitation pool that every mapping and cost record is actually
built on.  At the matched degree cap the surviving Majorana classes reach
exactly the determinants the excitation pool reaches, so the chain reconstructs
that pool rather than producing a smaller one.  The recorded verdict is ``not
falsified, but the content is pool-dependent'', the dependent work was retired
against it, and the record carries an explicit statement that nothing in it
licenses a resource claim of any kind.

This result shows why a filter must be evaluated on the pool used by the
solver. Removing redundant candidates from a wider pool does not establish a
saving against an already compact excitation pool.

\subsection{Noise-aware selection and independent checks}

A preregistered equal-shot simulation compared a channel-defect penalty with
unpenalized selection on spin and molecular cases. Both arms selected the same
candidate throughout the declared grid, so the stated benefit criterion failed.
The comparison uses fresh evaluation samples and a paired simulation, and
charges both selection and evaluation shots. Channel knowledge is assumed and
calibration, gate-level noise, device time and classical penalty cost in seconds
are outside the boundary. This negative finding does not rule out other
selectors or workloads. An independent reconstruction checks the recorded
experiment while sharing input definitions and numerical dependencies. A
separate occupation-sector audit checks the published H$_4$ selected-prefix
energies and word support, but does not replay adaptive selection or certify
Python heap bytes. The declarations, observations and checkers are available
in the source repository.

\section{Scope and limitations}
\label{sec:limits}

\paragraph{Instance scale.}  Every quantitative result here is on a small active
space, constructed from committed integral files
\cite{knowles1989fcidump,sun2018pyscf}.  All the priced and sampled banks are
eight-qubit; the widest instance entering any structural readout is a
twelve-qubit water active space, and it is excluded from the cross-instance
setting-count verdict for using a different cover heuristic.  Nothing here is a
scaling result, and the block-size trade in particular is explicitly
instance-dependent in the only two instances measured.

\paragraph{Device cards are illustrative.}  Each card carries a calibration
status field, and the two non-baseline cards declare themselves illustrative
project scenarios rather than vendor data.  Routing is a declared scalar
surrogate, not a compiled layout.  Every runtime here is logical accounting, and
none of it is a hardware measurement or a prediction about a named device.

\paragraph{Search uncertainty is heuristic.}  The accuracy-matched costs come
from Monte Carlo bracketing on a fixed shot grid with declared replica counts.
The records label that uncertainty heuristic, separately from the exact-tier
label on the oracle comparison, and this paper does not merge them.

\paragraph{The negative results are bounded by their own scope.}  The contextual
screen is one bank under one set of declared gates; it does not establish that
contextual restriction cannot help an operator-response solver in general, only
that no rung of this screen admitted all four arms.  The
mapping-versus-instance refinement establishes nothing in the converse
direction, by its own recorded asymmetry.  The residual closure is post hoc and
covers one allocation and one estimator on its own banks; the clique comparison
decides each of its instances separately and pools none of them.

\paragraph{Evidence validation has historical limits.} Device ledgers explicitly
represent inadmissibility, and storage reports state the allocations they cover.
\cqcRecordsUnlabelled{} committed JSON records lack an evidence declaration, and
\cqcRecordsAnnotated{} take theirs from digest-bound annotations rather than from
their own bytes. New or changed JSON records require the versioned contract, but
in-package result types still carry string labels their constructors do not
validate, and series files are not inspected row by row. A metadata validator
does not verify physical results or device calibration. Downstream code must
still honor each result's status and assumptions.

\paragraph{Simulation and interoperability limits.} Sparse operator support can
grow rapidly, and dense statevector methods retain exponential storage cost.
Sector restriction helps only when the selected operations preserve that sector;
stabilizer execution supports its Clifford subset. Optional bridges have
different supported operations and do not imply universal circuit portability.
No benchmark here establishes relative speed or capability against another
framework.

\section{Reproducibility}
\label{sec:reproducibility}

Every empirical quantity in the body and tables is generated from a committed
record or a live census of the source tree. The manuscript checker rejects
numerals outside a short allowlist of conceptual notation. It regenerates
table fragments and compares them byte for byte, catching hand edits that an
input hash alone would miss. A figure manifest binds each vector asset to its
generator, input digests, and checksum. Scratch figure renders are not compared
byte for byte because plotting and font builds can serialize the same paths
differently.

The checker also rejects number words beside the nouns the census counts ---
gates, producers, records, modules. Where the prose states a pattern that a
record must keep --- identical scientific trajectories across storage arms,
negative residual verdicts, or the direction of a clique comparison --- the
generator asserts it and refuses to emit contradictory fragments. The reported
clique confidence level and decision threshold are generated from the frozen
declaration, whose digest must match the executed record.

The layer census requires every package module to belong to exactly one layer;
the gate census requires every benchmark gate to have a declared class.
Evidence-language checks preserve statements about admissibility, the bias
floor, indeterminate verdicts, and the limited scope of the lifetime bound.
These checks detect structural and reporting drift; scientific correctness
still depends on the independent validations and assumptions described above.

\section{Use and distribution}
\label{sec:discussion}

The package is useful as a research toolkit in three settings: developing
operator and subspace methods, comparing estimators on shared measurement data,
and testing resource tradeoffs under explicit assumptions. A common model and
matrix-element bank reduce duplicated implementation across these tasks.
Prepared inputs make repeated solves cheaper to organize, while separate
reference validation keeps its cost visible. These are practical software
benefits even when a particular algorithmic comparison is inconclusive.

The source includes an Apache license, package metadata, runnable examples,
optional dependency groups, and versioned tests and benchmark records. At this
manuscript revision it remains an alpha research package. The source package
is already publicly available at
\url{https://github.com/gutama/clifford_qc}. This statement does not assert that
a public package-index release or an archival identifier already exists.
For a source checkout, installation instructions and examples are in
\texttt{README.md}, repeated molecular workflows in \texttt{molecular/PIPELINE.md}, and
benchmark environments in \texttt{REPRODUCING.md}.

Users should identify the source version and dependency environment used in an
experiment. The library API is broader than the command-line pipeline, and
research interfaces can still change. A stable archival release would benefit
from consistently validated labels across in-package result types, a documented
compatibility policy, and a persistent citation for the released source. These
are distribution and maintenance needs, not evidence that the implemented
algorithms scale to larger problems.

\section{Conclusion}
\label{sec:conclusion}

\pkg{} connects sparse operator algebra, quantum models, eigensolvers, and
measurement analysis through reusable Python interfaces. Its useful features
include virtual operator-response bases, shared measurement caches, exact
residual diagnostics, independent preparation and validation, and configurable
coefficient storage whose memory--time trade is now measured. The benchmark
records demonstrate how these components support controlled comparisons and
identify accuracy or cost assumptions that fail. Small-system evidence,
illustrative device cards, post-hoc designs where they are declared, and
in-package result labels that are not yet validated limit the claims. Within
those limits, the package provides a working foundation for reproducible
quantum simulation research.

\section*{Code and data availability}

The package source, the committed records behind every number in this paper,
the generators that turn those records into its tables and figures, and the
checker that gates the manuscript against them are versioned together in one
Apache-licensed repository at
\url{https://github.com/gutama/clifford_qc}. The manuscript sources are under
\texttt{paper\_architecture/}; \texttt{REPRODUCING.md} records the environment
each record was produced in, and \texttt{CITATION.cff} carries the citation
metadata for the source itself. The source revision corresponding to this
manuscript contains the required data; rebuilding also requires the documented
software environment and TeX dependencies. Public source access does not imply a package-index release or a persistent
archival identifier; users should record the source revision they used.

\appendix
\section{Evidence and validation details}
\label{app:evidence}

\subsection{Evidence coverage}

\begin{table}
  \caption{How the committed record set declares the evidence its numbers
  carry.  Counts are recomputed from
  \texttt{benchmarks/reference\_results/} at manuscript build time.}
  \label{tab:evidence}
  \scriptsize
  \begin{ruledtabular}
  \begin{tabular}{@{}lrl@{}}
  Declaration form & Records & Example \\
  \colrule
  \input{tables/evidence_coverage.tex}
  \end{tabular}
  \end{ruledtabular}
\end{table}

Table~\ref{tab:evidence} measures the current coverage. Of \cqcJsonRecords{} committed JSON records,
\cqcRecordsLabelled{} declare a single top-level evidence tier, drawn from
\cqcDistinctTiers{} distinct tier strings; \cqcRecordsPerQuantity{} declare a
per-quantity mapping, either because a record reports quantities at different
tiers or because the versioned contract below asks every new record for one;
\cqcRecordsNested{} nest the tier inside a
sub-object; \cqcRecordsRoleWithNested{} declare an evidence \emph{role} at the
top level with tiers inside sub-objects; \cqcRecordsRoleOnly{} carry a role with
no tier at all; and \cqcRecordsUnlabelled{} carry no record-level evidence field
whatsoever.  A further \cqcSeriesFiles{} non-JSON series files --- line-delimited
JSON, CSV and Markdown summaries --- are counted but not inspected for per-row
declarations by this census.

The selection vocabulary is a closed enumeration in code. Historical result
types retain their original string labels and constructors; they are not
retroactively converted to a new vocabulary. A versioned evidence contract
requires every quantity a record reports to declare one tier and a nonempty
scope. It is validated at the opt-in producer stamp and by an automatic metadata
gate on every new or changed JSON record. Only the byte-identical historical set
is grandfathered, by a frozen digest manifest, and those records keep their
bytes: \cqcRecordsAnnotated{} of them take their declaration from a separately
disclosed annotation bound to the record's digest, which the census reads
without changing the original record or its execution identity. Scientific
values and the truth of each claim still require their independent record
gates; enumerating a tier does not establish it.

\section{Producer and gate census}
\label{app:gates}

The census identifies \cqcProducers{} producers and \cqcGates{} gates, of
which \cqcGatesInWorkflow{} appear in CI: \cqcGatesAutomatic{} run on every
pull request, \cqcGatesConditional{} are conditional, and \cqcGatesFullReplay{}
require explicit dispatch. The remaining \cqcGatesAbsent{} are not in CI.
The shared runner records command failures and receipts; affected restart
replays also run on a schedule. The census describes coverage, not correctness.

\begin{table*}
  \caption{What each class of benchmark gate asserts.  Counts are recomputed
  from \texttt{benchmarks/} at manuscript build time, and the generator refuses
  to emit the table unless every gate on disk has a declared class.}
  \label{tab:gates}
  \footnotesize
  \begin{ruledtabular}
  \begin{tabular}{lrp{0.62\textwidth}}
  Class & Gates & Asserts \\
  \colrule
  \input{tables/gate_classes.tex}
  \end{tabular}
  \end{ruledtabular}
\end{table*}

Among the producers, \cqcSameStemPairs{} have a same-stem numerical rebuild;
\cqcProducersWithoutGate{} do not. The validation-ladder record belongs to the
latter group; its historical cross-paper inputs are preserved rather than
regenerated. Table~\ref{tab:gates} distinguishes numerical, structural and
provenance checks.

The \cqcGatesPrereg{} preregistration checks verify configurations frozen before
sampling \cite{nosek2018preregistration}; \cqcGatesPostHoc{} check declarations
made post hoc, after earlier observations. The \cqcGatesReplay{} independent
replays check recorded trajectories rather than global optimality. The
\cqcGatesLineage{} lineage checks do not rebuild values, and
\cqcGatesCrossArtifact{} compare independently produced artifacts. These
categories carry different evidential weight.

\begin{acknowledgments}
We thank the maintainers of the open-source quantum software ecosystem that
\pkg{} bridges to.  The framework is a pre-release research codebase; the
records, generators, gates and tests that produced every number in this
manuscript are versioned together.
\end{acknowledgments}

\bibliographystyle{apsrev4-2}
\bibliography{references}

\end{document}

%% file: tables/numbers.tex
\newcommand{\cqcBankBasis}{\ensuremath{5}}
\newcommand{\cqcBankWords}{\ensuremath{1{,}814}}
\newcommand{\cqcBeHBias}{\ensuremath{0.0033}}
\newcommand{\cqcBeHCXRatio}{\ensuremath{1.15}}
\newcommand{\cqcBeHCXRatioWorst}{\ensuremath{2.79}}
\newcommand{\cqcBeHPrepReduction}{\ensuremath{96.0\%}}
\newcommand{\cqcBeHSettingsFull}{\ensuremath{14}}
\newcommand{\cqcBeHSettingsQWC}{\ensuremath{353}}
\newcommand{\cqcBothSaveMax}{\ensuremath{73\%}}
\newcommand{\cqcBothSaveMin}{\ensuremath{62\%}}
\newcommand{\cqcBridgeModules}{\ensuremath{6}}
\newcommand{\cqcCrossAgreeing}{\ensuremath{6}}
\newcommand{\cqcCrossChecked}{\ensuremath{8}}
\newcommand{\cqcDistinctTiers}{\ensuremath{5}}
\newcommand{\cqcFullCertifiedShots}{\ensuremath{2^{24}}}
\newcommand{\cqcGOneExcitation}{\ensuremath{0}}
\newcommand{\cqcGOneMajorana}{\ensuremath{522}}
\newcommand{\cqcGates}{\ensuremath{79}}
\newcommand{\cqcGatesAbsent}{\ensuremath{7}}
\newcommand{\cqcGatesAutomatic}{\ensuremath{61}}
\newcommand{\cqcGatesConditional}{\ensuremath{0}}
\newcommand{\cqcGatesCrossArtifact}{\ensuremath{5}}
\newcommand{\cqcGatesFullReplay}{\ensuremath{11}}
\newcommand{\cqcGatesInWorkflow}{\ensuremath{72}}
\newcommand{\cqcGatesLineage}{\ensuremath{3}}
\newcommand{\cqcGatesPostHoc}{\ensuremath{6}}
\newcommand{\cqcGatesPrereg}{\ensuremath{13}}
\newcommand{\cqcGatesReplay}{\ensuremath{11}}
\newcommand{\cqcHFourBias}{\ensuremath{3.019}}
\newcommand{\cqcHFourCXAvoided}{\ensuremath{2.22}}
\newcommand{\cqcHFourCXRatio}{\ensuremath{0.45}}
\newcommand{\cqcHFourPolicySpread}{\ensuremath{5\%}}
\newcommand{\cqcHFourPrepReduction}{\ensuremath{93.0\%}}
\newcommand{\cqcHFourSettingsFull}{\ensuremath{64}}
\newcommand{\cqcHFourSettingsQWC}{\ensuremath{913}}
\newcommand{\cqcIonRuntimeRatio}{\ensuremath{0.699}}
\newcommand{\cqcJsonRecords}{\ensuremath{56}}
\newcommand{\cqcKStarIon}{\ensuremath{2}}
\newcommand{\cqcKStarIonRegion}{1, 2, 4}
\newcommand{\cqcKStarLogical}{\ensuremath{4}}
\newcommand{\cqcKernelLines}{\ensuremath{2{,}264}}
\newcommand{\cqcLines}{\ensuremath{34{,}080}}
\newcommand{\cqcLogicalRuntimeRatio}{\ensuremath{0.113}}
\newcommand{\cqcMappingSpreadKOne}{\ensuremath{3.48}}
\newcommand{\cqcMeasurementLines}{\ensuremath{5{,}456}}
\newcommand{\cqcModules}{\ensuremath{126}}
\newcommand{\cqcPNineteenDelta}{\ensuremath{0.05}}
\newcommand{\cqcPNineteenLooks}{\ensuremath{17}}
\newcommand{\cqcPNineteenReductionFactor}{\ensuremath{2}}
\newcommand{\cqcPNineteenSystems}{\ensuremath{3}}
\newcommand{\cqcPNineteenTarget}{\ensuremath{1.6}}
\newcommand{\cqcPackSaveMax}{\ensuremath{63\%}}
\newcommand{\cqcPackSaveMin}{\ensuremath{49\%}}
\newcommand{\cqcPolicyArms}{\ensuremath{4}}
\newcommand{\cqcPolicySystems}{\ensuremath{5}}
\newcommand{\cqcProducers}{\ensuremath{76}}
\newcommand{\cqcProducersWithoutGate}{\ensuremath{24}}
\newcommand{\cqcQWCCertifiedShots}{\ensuremath{2^{29}}}
\newcommand{\cqcRFourAACaseBias}{\ensuremath{0.0695}}
\newcommand{\cqcRFourAACaseWords}{\ensuremath{1{,}223}}
\newcommand{\cqcRFourAAdmissibleBias}{\ensuremath{0.533}}
\newcommand{\cqcRFourACSACaseBias}{\ensuremath{5.90}}
\newcommand{\cqcRFourACSBias}{\ensuremath{5.35}}
\newcommand{\cqcRFourACeiling}{\ensuremath{2{,}048}}
\newcommand{\cqcRFourAFullWords}{\ensuremath{14{,}350}}
\newcommand{\cqcRFourAPassing}{\ensuremath{0}}
\newcommand{\cqcRFourARemovedFraction}{\ensuremath{15.3\%}}
\newcommand{\cqcRFourARungs}{\ensuremath{7}}
\newcommand{\cqcRThreeCCells}{\ensuremath{40}}
\newcommand{\cqcRThreeCCensored}{\ensuremath{2}}
\newcommand{\cqcRThreeCFailures}{\ensuremath{0}}
\newcommand{\cqcRThreeCFinite}{\ensuremath{38}}
\newcommand{\cqcRThreeCSCPriced}{\ensuremath{23}}
\newcommand{\cqcRThreeCSolves}{\ensuremath{19{,}300}}
\newcommand{\cqcRThreeDCells}{\ensuremath{7}}
\newcommand{\cqcRThreeDInstanceSupport}{\ensuremath{4.00}}
\newcommand{\cqcRThreeDMappingSupport}{\ensuremath{2.19}}
\newcommand{\cqcRecordsAnnotated}{\ensuremath{8}}
\newcommand{\cqcRecordsLabelled}{\ensuremath{17}}
\newcommand{\cqcRecordsNested}{\ensuremath{5}}
\newcommand{\cqcRecordsPerQuantity}{\ensuremath{32}}
\newcommand{\cqcRecordsRoleOnly}{\ensuremath{0}}
\newcommand{\cqcRecordsRoleWithNested}{\ensuremath{2}}
\newcommand{\cqcRecordsUnlabelled}{\ensuremath{0}}
\newcommand{\cqcResidualBanks}{\ensuremath{5}}
\newcommand{\cqcResidualBar}{\ensuremath{10}}
\newcommand{\cqcResidualFCAltMax}{\ensuremath{117{,}681}}
\newcommand{\cqcResidualFCAltMin}{\ensuremath{14.6}}
\newcommand{\cqcResidualFCBanks}{\ensuremath{4}}
\newcommand{\cqcResidualFCMax}{\ensuremath{26.7}}
\newcommand{\cqcResidualFCMin}{\ensuremath{12.3}}
\newcommand{\cqcResidualQWCBanks}{\ensuremath{5}}
\newcommand{\cqcResidualQWCMax}{\ensuremath{184{,}358}}
\newcommand{\cqcResidualQWCMin}{\ensuremath{21.0}}
\newcommand{\cqcRitzSupport}{\ensuremath{966}}
\newcommand{\cqcSCFidelityFloor}{\ensuremath{0.50}}
\newcommand{\cqcSCInadmissibleSettings}{\ensuremath{12}}
\newcommand{\cqcSCMeanFidelity}{\ensuremath{0.452}}
\newcommand{\cqcSCSettings}{\ensuremath{14}}
\newcommand{\cqcSameStemPairs}{\ensuremath{52}}
\newcommand{\cqcSeriesFiles}{\ensuremath{21}}
\newcommand{\cqcShotRatio}{\ensuremath{0.03125}}
\newcommand{\cqcShotRatioInverse}{\ensuremath{32}}
\newcommand{\cqcStreamSaveMax}{\ensuremath{68\%}}
\newcommand{\cqcStreamSaveMin}{\ensuremath{63\%}}
\newcommand{\cqcStructuralInstance}{\ensuremath{3.98}}
\newcommand{\cqcStructuralMapping}{\ensuremath{13.07}}
\newcommand{\cqcStructuralMargin}{\ensuremath{0.305}}
\newcommand{\cqcSubspaceLines}{\ensuremath{15{,}467}}
\newcommand{\cqcTargetMilliHartree}{\ensuremath{1.6}}
\newcommand{\cqcTestLines}{\ensuremath{44{,}405}}
\newcommand{\cqcTestModules}{\ensuremath{187}}

%% file: tables/layers.tex
Algebra kernel & \ensuremath{15} & \ensuremath{2{,}264} & \ensuremath{6.6\%} \\
Program IR & \ensuremath{4} & \ensuremath{1{,}246} & \ensuremath{3.7\%} \\
Execution & \ensuremath{10} & \ensuremath{2{,}046} & \ensuremath{6.0\%} \\
Measurement & \ensuremath{17} & \ensuremath{5{,}456} & \ensuremath{16.0\%} \\
Problem definition & \ensuremath{9} & \ensuremath{2{,}739} & \ensuremath{8.0\%} \\
Algorithms & \ensuremath{10} & \ensuremath{1{,}967} & \ensuremath{5.8\%} \\
Subspace solvers & \ensuremath{46} & \ensuremath{15{,}467} & \ensuremath{45.4\%} \\
Cross-layer orchestration & \ensuremath{3} & \ensuremath{835} & \ensuremath{2.5\%} \\
Evidence and reproduction & \ensuremath{5} & \ensuremath{1{,}335} & \ensuremath{3.9\%} \\
Optional bridges & \ensuremath{6} & \ensuremath{687} & \ensuremath{2.0\%} \\
Package surface & \ensuremath{1} & \ensuremath{38} & \ensuremath{0.1\%} \\
\colrule
Package total & \ensuremath{126} & \ensuremath{34{,}080} & \ensuremath{100.0\%} \\
Test suite & \ensuremath{187} & \ensuremath{44{,}405} & --- \\

%% file: tables/device_cards.tex
logical-alltoall & \ensuremath{0} & \ensuremath{0.05} & \ensuremath{0.3} & \ensuremath{1} & \ensuremath{1} & \ensuremath{0} & \ensuremath{0} & \ensuremath{0} & \ensuremath{1} & \ensuremath{1} & \ensuremath{0.5} \\
ion-like & \ensuremath{50} & \ensuremath{10} & \ensuremath{200} & \ensuremath{100} & \ensuremath{50} & \ensuremath{0.0005} & \ensuremath{0.005} & \ensuremath{0.01} & \ensuremath{1} & \ensuremath{1} & \ensuremath{0.5} \\
superconducting-like & \ensuremath{1} & \ensuremath{0.05} & \ensuremath{0.3} & \ensuremath{1} & \ensuremath{1} & \ensuremath{0.001} & \ensuremath{0.01} & \ensuremath{0.02} & \ensuremath{3} & \ensuremath{2} & \ensuremath{0.5} \\

%% file: tables/storage_policy.tex
H$_4$ & \ensuremath{11} & \ensuremath{0.21} & \ensuremath{0.21} & \ensuremath{0.21} & \ensuremath{0.20} & \ensuremath{0.04} & \ensuremath{0.05} & \ensuremath{0.05} & \ensuremath{0.05} \\
BeH$_2$ & \ensuremath{11} & \ensuremath{4.54} & \ensuremath{2.32} & \ensuremath{1.66} & \ensuremath{1.71} & \ensuremath{3.31} & \ensuremath{5.04} & \ensuremath{4.81} & \ensuremath{5.99} \\
BeH$_2$, stretched & \ensuremath{31} & \ensuremath{11.12} & \ensuremath{4.26} & \ensuremath{3.55} & \ensuremath{3.21} & \ensuremath{12.74} & \ensuremath{21.92} & \ensuremath{18.61} & \ensuremath{22.74} \\
H$_2$O & \ensuremath{21} & \ensuremath{7.90} & \ensuremath{3.64} & \ensuremath{2.56} & \ensuremath{2.31} & \ensuremath{9.19} & \ensuremath{15.89} & \ensuremath{13.53} & \ensuremath{17.72} \\
H$_2$O, stretched & \ensuremath{31} & \ensuremath{9.87} & \ensuremath{3.67} & \ensuremath{3.20} & \ensuremath{2.63} & \ensuremath{14.39} & \ensuremath{25.31} & \ensuremath{20.89} & \ensuremath{26.15} \\

%% file: tables/hierarchy.tex
H$_4$ & \ensuremath{1} & \ensuremath{913} & \ensuremath{0.0} & \ensuremath{0} & \ensuremath{7{,}304{,}000} & --- \\
 & \ensuremath{2} & \ensuremath{647} & \ensuremath{4.3} & \ensuremath{4} & \ensuremath{5{,}176{,}000} & \ensuremath{0.09} \\
 & \ensuremath{4} & \ensuremath{238} & \ensuremath{13.8} & \ensuremath{14} & \ensuremath{1{,}904{,}000} & \ensuremath{0.21} \\
 & \ensuremath{8} & \ensuremath{64} & \ensuremath{29.5} & \ensuremath{35} & \ensuremath{512{,}000} & \ensuremath{0.45} \\
\colrule
BeH$_2$ & \ensuremath{1} & \ensuremath{353} & \ensuremath{0.0} & \ensuremath{0} & \ensuremath{2{,}824{,}000} & --- \\
 & \ensuremath{2} & \ensuremath{41} & \ensuremath{2.7} & \ensuremath{1} & \ensuremath{328{,}000} & \ensuremath{2.79} \\
 & \ensuremath{4} & \ensuremath{26} & \ensuremath{7.4} & \ensuremath{10} & \ensuremath{208{,}000} & \ensuremath{1.69} \\
 & \ensuremath{8} & \ensuremath{14} & \ensuremath{21.1} & \ensuremath{23} & \ensuremath{112{,}000} & \ensuremath{1.15} \\

%% file: tables/phase14b.tex
Compiled settings & \ensuremath{353} & \ensuremath{14} & \ensuremath{0.040} \\
Settings the Ritz functional touches & \ensuremath{179} & \ensuremath{7} & \ensuremath{0.039} \\
Readable word--setting pairs & \ensuremath{4{,}790} & \ensuremath{3{,}284} & \ensuremath{0.686} \\
One-qubit gates, summed over settings & \ensuremath{3{,}648} & \ensuremath{297} & \ensuremath{0.081} \\
Two-qubit gates, summed over settings & \ensuremath{0} & \ensuremath{296} & --- \\
Certified total physical shots & \ensuremath{2^{29}} & \ensuremath{2^{24}} & \ensuremath{0.03125} \\
\colrule
$C_{\mathrm{time}}(\epsilon)$, ion-like (s) & \ensuremath{134{,}835} & \ensuremath{94{,}291} & \ensuremath{0.699} \\
$C_{\mathrm{time}}(\epsilon)$, logical-alltoall (s) & \ensuremath{1{,}108} & \ensuremath{125} & \ensuremath{0.113} \\
$C_{\mathrm{time}}(\epsilon)$, superconducting-like (s) & \ensuremath{2{,}294} & inadmissible & --- \\
Settings below the fidelity floor, superconducting-like & \ensuremath{0} & \ensuremath{12} & --- \\

%% file: tables/phase19.tex
H$_4$ & QWC & \ensuremath{68} & \ensuremath{0} & $>$\ensuremath{2^{32}} & censored & censored & censored & material reduction, censored baseline \\
 & clique & \ensuremath{45} & \ensuremath{1{,}236} & \ensuremath{2^{31}} & \ensuremath{25{,}373} & \ensuremath{23{,}713{,}093} & inadmissible &  \\
\colrule
LiH & QWC & \ensuremath{42} & \ensuremath{0} & \ensuremath{2^{29}} & \ensuremath{1{,}112} & \ensuremath{135{,}709} & \ensuremath{2{,}296} & no material reduction \\
 & clique & \ensuremath{54} & \ensuremath{906} & \ensuremath{2^{30}} & \ensuremath{9{,}499} & \ensuremath{7{,}519{,}536} & inadmissible &  \\
\colrule
BeH$_2$ & QWC & \ensuremath{9} & \ensuremath{0} & \ensuremath{2^{26}} & \ensuremath{135} & \ensuremath{15{,}915} & \ensuremath{279} & material reduction \\
 & clique & \ensuremath{36} & \ensuremath{166} & \ensuremath{2^{25}} & \ensuremath{121} & \ensuremath{53{,}531} & \ensuremath{521} &  \\

%% file: tables/cost.tex
ion-like & single assignment & \ensuremath{2} & 1, 2, 4 & region & \ensuremath{364} & \texttt{bk+2q} & \ensuremath{1.74} \\
ion-like & pooled & \ensuremath{2} & 1, 2 & region & \ensuremath{91.1} & \texttt{bk+2q} & \ensuremath{1.74} \\
logical-alltoall & single assignment & \ensuremath{4} & 1, 2, 4, 8 & region & \ensuremath{1.55} & \texttt{bk} & \ensuremath{3.87} \\
logical-alltoall & pooled & \ensuremath{2} & 1, 2, 4, 8 & region & \ensuremath{0.425} & \texttt{bk+2q} & \ensuremath{1.62} \\
superconducting-like & single assignment & \ensuremath{2} & 1, 2, 4 & region & \ensuremath{4.33} & \texttt{bk+2q} & \ensuremath{1.84} \\
superconducting-like & pooled & \ensuremath{2} & 1, 2 & region & \ensuremath{1.08} & \texttt{bk+2q} & \ensuremath{1.84} \\
\colrule
\multicolumn{8}{p{0.96\textwidth}}{Rungs the superconducting-like card refuses to price: $k={8}$. Accuracy target \ensuremath{1.6} mHa; H$_4$ is not priced at all, its own subspace bias being \ensuremath{3.019} mHa.} \\

%% file: tables/qr3.tex
QWC settings, matched greedy & structural & widest mapping vs.\ narrowest instance & \ensuremath{13.07} & \ensuremath{3.98} & mapping spread not smaller than instance spread \\
Mean Pauli word weight & structural & widest mapping vs.\ narrowest instance & \ensuremath{1.57} & \ensuremath{1.49} & mapping spread not smaller than instance spread \\
\colrule
Accuracy-matched $C_{\mathrm{time}}(\epsilon)$ & exact & point estimate on the supporting cells & \ensuremath{17.52} & \ensuremath{1.02} & indeterminate at this shot grid \\
Accuracy-matched $C_{\mathrm{time}}(\epsilon)$ & exact & conservative support of the same cells & \ensuremath{1.36} & \ensuremath{4.08} & indeterminate at this shot grid \\
R3d midpoint refinement & exact & support after added endpoints & \ensuremath{2.19} & \ensuremath{4.00} & indeterminate after refinement \\

%% file: tables/r4a.tex
\ensuremath{1} & \ensuremath{7} & \ensuremath{5.35} & \ensuremath{3{,}864} & \ensuremath{5.90} & \ensuremath{40} & \ensuremath{15.3\%} & \ensuremath{1} \\
\ensuremath{2} & \ensuremath{6} & \ensuremath{5.35} & \ensuremath{997} & \ensuremath{5.90} & \ensuremath{33} & \ensuremath{15.3\%} & \ensuremath{1} \\
\ensuremath{3} & \ensuremath{5} & \ensuremath{5.90} & \ensuremath{124} & \ensuremath{5.90} & \ensuremath{19} & \ensuremath{15.4\%} & \ensuremath{1} \\
\ensuremath{4} & \ensuremath{4} & \ensuremath{5.90} & \ensuremath{14} & \ensuremath{5.90} & \ensuremath{14} & \ensuremath{15.4\%} & \ensuremath{1} \\
\ensuremath{5} & \ensuremath{3} & \ensuremath{5.90} & \ensuremath{6} & \ensuremath{5.90} & \ensuremath{6} & \ensuremath{15.5\%} & \ensuremath{1} \\
\ensuremath{6} & \ensuremath{2} & \ensuremath{5.90} & \ensuremath{3} & \ensuremath{5.90} & \ensuremath{3} & \ensuremath{15.5\%} & \ensuremath{1} \\
\ensuremath{7} & \ensuremath{1} & \ensuremath{5.90} & \ensuremath{1} & \ensuremath{5.90} & \ensuremath{1} & \ensuremath{15.5\%} & \ensuremath{1} \\
\colrule
\multicolumn{8}{p{0.96\textwidth}}{Unrestricted arms, identical at every rung: \texttt{full\_qse} bias \ensuremath{0.0033} mHa, word universe \ensuremath{14{,}350} (ceiling \ensuremath{2{,}048}); \texttt{acase} bias \ensuremath{0.0695} mHa, word universe \ensuremath{1{,}223}. Admissible bias \ensuremath{0.533} mHa.} \\

%% file: tables/evidence_coverage.tex
Top-level tier & \ensuremath{17} & \texttt{bank-storage-ledger} \\
Per-quantity mapping & \ensuremath{32} & \texttt{clifford-hierarchy-beh2-v2} \\
Role plus nested tier & \ensuremath{2} & \texttt{qr3b-instance-preflight} \\
Role, not a tier & \ensuremath{0} & --- \\
Tier basis, not a tier & \ensuremath{0} & --- \\
Nested in sub-object & \ensuremath{5} & \texttt{bank-policy-matrix} \\
No evidence & \ensuremath{0} & --- \\
\colrule
JSON records & \ensuremath{56} & \ensuremath{5} tier strings \\
Series files (JSONL, CSV, MD) & \ensuremath{21} & not inspected here \\

%% file: tables/gate_classes.tex
Kernel fixture audit & \ensuremath{1} & rebuilds named blade decisions and support bounds; audits measured timings without speed thresholds \\
Exact trajectory replay & \ensuremath{11} & independently replays occupation states, root pencils, noise or optimizer calls and counters; not global optimality \\
Observed resource equivalence & \ensuremath{1} & derives accounting and equivalence from isolated cells; optionally rebuilds final pencils \\
Sampled inference accounting & \ensuremath{2} & checks committed declaration/execution bindings, every observed attempt, physical costs and unconditional target attainment; full shot replay is explicit \\
Value-rebuilt & \ensuremath{34} & recomputes its same-stem producer's record and fails on numeric drift \\
Preregistration & \ensuremath{13} & validates a frozen plan without outcomes; may rebuild structural inputs \\
Post-hoc declaration & \ensuremath{6} & freezes a disclosed post-hoc design; commit order protects identity, not preregistration \\
Lineage and environment & \ensuremath{3} & provenance stamping, environment migration, regeneration identity \\
Plan ledger & \ensuremath{1} & the phase status file against the plan document and the source tree \\
Documentation & \ensuremath{1} & every documented command, flag, constant, and record still exists \\
Evidence metadata & \ensuremath{1} & validates versioned per-quantity labels and digest-bound historical annotations; not scientific values \\
Cross-artifact & \ensuremath{5} & compares separately produced repository artifacts rather than a same-stem rebuild \\
\colrule
Gates, all classes & \ensuremath{79} & \ensuremath{72} named in the CI workflow \\
Run on every pull request & \ensuremath{61} & \ensuremath{0} selected by affected paths; \ensuremath{11} in scheduled or explicit full replay \\
Producers & \ensuremath{76} & \ensuremath{52} have a same-stem gate \\
Producers with no value gate & \ensuremath{24} & manuscript evidence, drivers, exploratory output \\